%% file: main.tex
\def \<{\langle}
\def \>{\rangle}

\documentclass[a4paper,11pt]{article}  

\usepackage{jcappub,graphicx,epsfig,natbib,color,times,bm,amsmath,multirow,hyperref, booktabs}

\usepackage{float}
\usepackage{subfig}
\usepackage[flushleft]{threeparttable}

\newcommand{\GeV}{\text{GeV}}

\newcommand{\vev}[1]{\left\langle #1 \right\rangle}

\newcommand{\be}{\begin{equation}}
\newcommand{\ee}{\end{equation}}
\newcommand{\bea}{\begin{eqnarray}}
\newcommand{\eea}{\end{eqnarray}}

\newcommand{\ckk}[1]{\textcolor{red}{#1}}

\DeclareRobustCommand{\Eq}[1]{Eq.~(\ref{#1})}

\DeclareRobustCommand{\r}[1]{{\rm #1}}

\newcommand{\rd}{\r{d}}
\newcommand{\inj}{\left[\frac{\rd E}{\rd V \rd t}\right]_{\r{inj}}}
\newcommand{\dep}{\left[\frac{\rd E}{\rd V \rd t}\right]_{\r{dep,\r{c}}}}
\newcommand{\sv}{\left<\sigma v\right>}
\newcommand{\pann}{\sv / m_{\chi}}
\newcommand{\fc}{f_{\r{c}}}
\newcommand{\xe}{x_{\r{e}}}
\newcommand{\Tk}{T_{\r{k}}}
\newcommand{\alm}{a_{\ell m}}

\title{Dark matter search with CMB: a study of foregrounds}

\input{authors.tex}

\begin{document}

\abstract{
\input{abstract.tex}

}

\maketitle

\section{Introduction}\label{sec:intro}
\input{introduction.tex}

\section{Indirect dark matter search with CMB}\label{sec:dm_effects}
\input{effect_dm.tex}


\section{Data simulation and analysis}\label{sec:simu}
\input{simulation_analysis.tex}

\section{Results}\label{sec:result}
\input{results.tex}

\section{Discussions}\label{sec:conclusions}
\input{conclusion.tex}

\acknowledgments
Authors thank Hua Zhai, Yi-Wen Wu and members of the AliCPT group for helpful discussions. This work is supported by the NSFC under grant No.11653002 and No. 12275278, and in part by the National Key R\&D Program of China No.2020YFC2201600. 

\nocite{*}
\bibliography{main}
\bibliographystyle{JHEP} 

\end{document}

%% file: authors.tex
\author[a,b]{Zi-Xuan Zhang}\emailAdd{zhangzixuan@ihep.ac.cn}
\author[a,b]{Yi-Ming Wang}\emailAdd{wangyiming@ihep.ac.cn}
\author[a,b,c]{Junsong Cang}\emailAdd{cangjs@ihep.ac.cn}
\author[d]{Zirui Zhang}\emailAdd{zhzr@mail.sdu.edu.cn}
\author[a]{Yang Liu}\emailAdd{liuy92@ihep.ac.cn}
\author[a]{Si-Yu Li}\emailAdd{lisy@ihep.ac.cn}
\author[a]{Yu Gao\footnote{Corresponding authors\label{corresauth}}}\emailAdd{gaoyu@ihep.ac.cn}
\author[a]{Hong Li\textsuperscript{\ref{corresauth}}}\emailAdd{hongli@ihep.ac.cn}

\affiliation[a]{Key Laboratory of Particle Astrophysics, Institute of High Energy Physics, Chinese Academy of Sciences, Beijing 100049, China}
\affiliation[b]{School of Physical Sciences, University of Chinese Academy of Sciences, Beijing, 100049, China}
\affiliation[c]{School of Physics, Henan Normal University, Xinxiang, China}
\affiliation[d]{Institute of Frontier and Interdisciplinary Science and Key Laboratory of Particle Physics and Particle Irradiation (MOE), Shandong University, Qingdao}

%% file: abstract.tex
The energy injected from dark matter annihilation and decay processes potentially raises the ionisation of the intergalactic medium and leaves visible footprints on the anisotropy maps of the cosmic microwave background (CMB). Galactic foregrounds emission in the microwave bands contaminate the CMB measurement and may affect the search for dark matter's signature.
In this paper, we construct a full CMB data and foreground simulation based on the design of the next-generation ground-based CMB experiments. The foreground residual after the components separation on maps is fully considered in our data analysis, accounting for various contamination from the emission of synchrotron, thermal dust, free-free and spinning dust.
We analyse the corresponding sensitivity on dark matter parameters from the temperature and polarization maps, and we find that the CMB foregrounds leave a non-zero yet controllable impact on the sensitivity. Comparing with statistics-only analysis, the CMB foreground residual leads to a factor of at most 19\% weakening on energy-injection constraints, depending on the specific dark matter process and experimental configuration. Strong limits on dark matter annihilation rate and decay lifetime can be expected after foreground subtraction.

%% file: introduction.tex
The existence of dark matter (DM) is supported by a wealth of astronomical observations, 
most notably the galactic rotation curves~\cite{Rubin:1970zza,Rubin:1980zd,vanAlbada:1984js} 
and bullet galaxy clusters~\cite{Clowe_2006,Clowe_2004,Markevitch_2002}.
The latest cosmological observations, 
such as the Planck satellite's measurement of the cosmic microwave background (CMB)~\cite{Planck:2018vyg} suggested that DM accounts for around 85$\%$ of the matter density of the Universe.
After decades of dedicated searches,
the particle nature of DM still remains elusive~\cite{Albouy:2022cin}.
A popular hypothesis predicts that DM can be a new particle beyond the standard model of particle physics. Highly motivated candidates include weakly interacting massive particles (WIMPs)~\cite{PhysRevLett.50.1419,PhysRevLett.89.211301,SERVANT2003391,ELLIS1984453},
sterile neutrinos~\cite{Bertone:2018krk,Shi:1998km,Laine:2008pg}, axions~\cite{Peccei:1977hh,Sikivie:1982qv,Preskill:1982cy} and many others.

Astrophysical indirect searches for particle DM look for the Standard Model (SM) products from DM annihilation and decay processes. 
If produced at high redshifts, 
these SM particles were much hotter than the temperature of the Universe and would deposit their energy into the cosmic environment~\cite{Padmanabhan:2005es}.
Such energy injections during the dark age of the Universe can ionise the neutral gas and increase the Compton scattering between CMB photons and free electrons, which can leave visible imprints on CMB anisotropy spectrum~\cite{Zhang:2007zzh,Madhavacheril:2013cna, Slatyer:2015kla}. 
Compared to high energy cosmic-ray searches 
,
the limits derived from CMB is more sensitive to energy injection around recombination~\cite{PhysRevD.85.043522}.
The CMB limits are comparably less dependent on WIMP model details, 
cover a very large DM mass above KeV, 
and serve an excellent complementary indirect search.
Currently the {\it Planck} data already set a stringent constraint\cite{Planck:2018vyg}, 
where a lower mass limit of thermal WIMPs is placed around the same order as in the results from Fermi-LAT~\cite{Fermi-LAT:2011vow}.
In the near future, 
a number of upcoming CMB missions\cite{CMB-S4:2016ple,Abazajian:2019eic,SimonsObservatory:2019qwx,Li:2017drr,Li:2018rwc} are expected to enter operation and deliver higher precision measurements, 
these dark matter limits can be expected to improve significantly~\cite{Cang:2020exa}.

As CMB experiments reach higher sensitivity, it becomes necessary to better understand and remove the contribution from foreground emission. 
Several well productive techniques are developed and provided in the literature for foreground subtraction. For example, the template removal methods in which foreground template are built based on the existing observations, 
such as template fitting of SEVEM (Spectral Estimation Via Expectation Maximisation)\cite{2003MNRAS.345.1101M, 2008A&A...491..597L, 2012MNRAS.420.2162F} and Delta-Map\cite{2019PTEP.2019c3E01I} methods, the parametric methods basing on the foreground models, like Commander\cite{2006ApJ...641..665E} and ForeGroundBuster\cite{fgb-Stompor:2008sf}.
On the other hand, a huge number of blind subtraction techniques advanced rapidly in recent years\cite{2019MNRAS.484.1616Z, 2013MNRAS.435...18B}, in which it does not rely on a prior knowledge of the foreground, but rather on the fact that the foreground and the CMB have different frequency dependencies and their different spatial distributions to achieve a separation between the two. 
While the data-driven blind subtraction method does not rely on prior information of the foreground, it inevitably suffers from foreground residuals. 
Also, as noise levels progressively decrease with the increasing size of modern detector arrays and the development in cryogenic readout, foreground residual will potentially become a non-negligible bottleneck in precision measurements~\cite{Abazajian:2019eic}. 
Thus, 
new-generation experiments demand adequate assessment of the foreground residual effects. 


In this paper, we carry out end-to-end simulations and data analysis for the next generation of ground-based CMB observations. In order to more fully assess the impact from the foreground, we simulated two different foreground models and adopted two different foreground removal methods, comparing their results and study the impact of foreground effects.
We use the Internal Linear Combination (ILC) method \cite{2003ApJS..148...97B},
which is one of the most efficient and widely adopted methods for component separation of CMB observation\cite{2013ApJS..208...20B, 2016A&A...594A...9P},
to separate the foreground emission from simulated maps to assess the impact of foreground residuals, 
and then explore the expected sensitivity to WIMP dark matter parameter space.

The paper is organized as follows: 
Section~\ref{sec:dm_effects} reviews the effects of dark matter on CMB,
Section~\ref{sec:simu} describes the multi-frequency map simulations,
our results are shown in Section~\ref{sec:result} and we conclude in Section~\ref{sec:conclusions}.

%% file: effect_dm.tex
The presence of DM decay and annihilation processes act as extra energy injection sources,
with the relevant energy injection rate per unit volume given by~\cite{Liu:2016cnk}

\be
\inj
=
\left\{
\begin{array}{l}
\Gamma_{\chi}\Omega_{\r{dm}} \rho_{\rm{cr}} (1+z)^3\ \ \ \ \ \ \ \ \ \ \rm{decay}
\\
\\
\frac{\sv}{m_{\chi}}\Omega_{\r{dm}}^2 \rho_{\rm{cr}}^2 (1+z)^6\ \ \ \  \rm{annihilation}
\end{array}
\right.
\label{eqsae24a}
\ee
where the subscript $\chi$ indicates the DM particle,
$\Gamma_{\chi}$ is the DM decay width,
defined as the inverse of DM decay lifetime $\tau_{\chi}$.
$\Omega_{\r{dm}}$ and $\rho_{\rm{cr}}$ are the current DM density parameter and critical density respectively,
$\sv$ is the thermally averaged DM annihilation cross-section,
$m_{\chi}$ is DM particle mass.
Note that in the second line of \Eq{eqsae24a} we have assumed homogeneous DM distribution.
At lower redshifts ($z \lesssim 50$) as the structure formation begins,
the concentration of DM particles inside DM halos will lead to a boost to the DM annihilation rate and thereby increasing the energy injection rate~\cite{Liu:2016cnk}.
However in~\cite{Cang:2020exa} it was shown that CMB is mostly unaffected by this effect because it is relatively insensitive to energy injection at low redshifts,
thus in what follows we will ignore DM clustering and assume homogeneous DM distribution.

The energy injected by DM is absorbed by the gas mainly through three channels:
ionisation (HIon) and excitation (Ly$\alpha$) of neutral hydrogen and the heating (Heat) of the gas~\cite{Slatyer:2015kla}.
DM can generally decay or annihilate into a number of standard model (SM) particles that further hadronize and decay, and eventually end up with stable particles such as $\r{e}^{\pm}$, $\nu$ and $\gamma$.
Since $\r{e}^{\pm}$ and $\gamma$ make up the majority of the end-state particles that interact with gas,
one can calculate the energy deposited into different channels by tracking the energy loss process of $\r{e}^{\pm}$ and $\gamma$,
which can be described by a transfer function $\mathcal{T}^{\alpha}_{\r{c}}(z,E,z')$.
For a particle $\alpha \in [\r{e}^{\pm}, \gamma]$ injected at redshift $z'$ with energy $E$,
$\mathcal{T}^{\alpha}_{\r{c}}(z,E,z')$ describes the fraction of $E$ deposited into channel c during unit $\r{ln}(1+z)^{-1}$ interval near $z$.
In this work we adopt the $\mathcal{T}^{\alpha}_{\r{c}}(z,E,z')$ function provided in ~\cite{Slatyer:2015kla},
and it can be shown analytically that the energy deposition rate per unit volume $\left[{{\rm{d}}E}/{{\rm{d}}V{\rm{d}}t}\right]_{\rm{dep,c}}$ (hereafter dubbed deposition rate for simplicity) is given by~\cite{Cang:2021owu}
\be
\dep
(z)
=
(1+z)^3 H(z)
\int
\frac
{\r{d} z'}
{(1+z')^4 H(z')}
\sum_{\alpha = \gamma, e^{\pm}}
\left[
\int \r{d}E
\ 
E
\mathcal{T}^{\alpha}_{\r{c}}(z,E,z')
\mathcal{I}^{\alpha}(E,z')
\right]
,
\label{Dep_EFF_EQ}
\ee
here the subscript dep indicates the deposition rate,
$\r{c} \in [\r{HIon, Ly \alpha, Heat}]$ represents the deposition channel,
$\mathcal{I}^{\alpha} \equiv \rd N^{\alpha}/ \rd E \rd V \rd t$ is the number of particle $\alpha$ injected per unit time, volume and energy.

For simplicity,
here we will consider DM annihilation/decay into $e^{\pm}$ and $\gamma \gamma$ channels,
for which $\mathcal{I}^{\alpha}$ is given by,
\be
\mathcal{I}^{\alpha}
=
\frac{a}{E'}
\inj
\delta_{\r{D}}(E-E'),
\label{saigdyuwg7}
\ee
here the factor $a$ equals $1/2$ and 1 for $e^{\pm}$ and $\gamma \gamma$ channels respectively,
$E'$ equals $m_{\chi}/2$ for decay and $m_{\chi}$ for annihilation,
$\delta_{\r{D}}$ is the Dirac delta function.
Inserting \Eq{saigdyuwg7} into \Eq{Dep_EFF_EQ},
one finds that
\be
\dep
=
\fc
\times
\inj
,
\label{78he3jgddfsay_}
\ee
where the deposition efficiency $\fc$ describes the fraction of injected energy deposited into channel c,
\be
\fc^{\r{dec}}
=
\frac{H(z)}{2}
\int
\frac
{\r{d} z'}
{(1+z') H(z')}
\mathcal{T}^{\alpha}_{\r{c}}(z,m_{\chi}/2,z')
,
\label{sadwgs37tgyyy}
\ee

\be
\fc^{\r{ann}}
=
\frac{H(z)}{(1+z)^3}
\int
\frac
{\r{d} z'}
{(1+z')^{-2} H(z')}
\mathcal{T}^{\alpha}_{\r{c}}(z,m_{\chi},z')
,
\label{sadwgs37tgyaa}
\ee
and the superscripts dec and ann denote decay and annihilation respectively.

\begin{figure}[t]
    \centering
    \includegraphics[width=\textwidth]{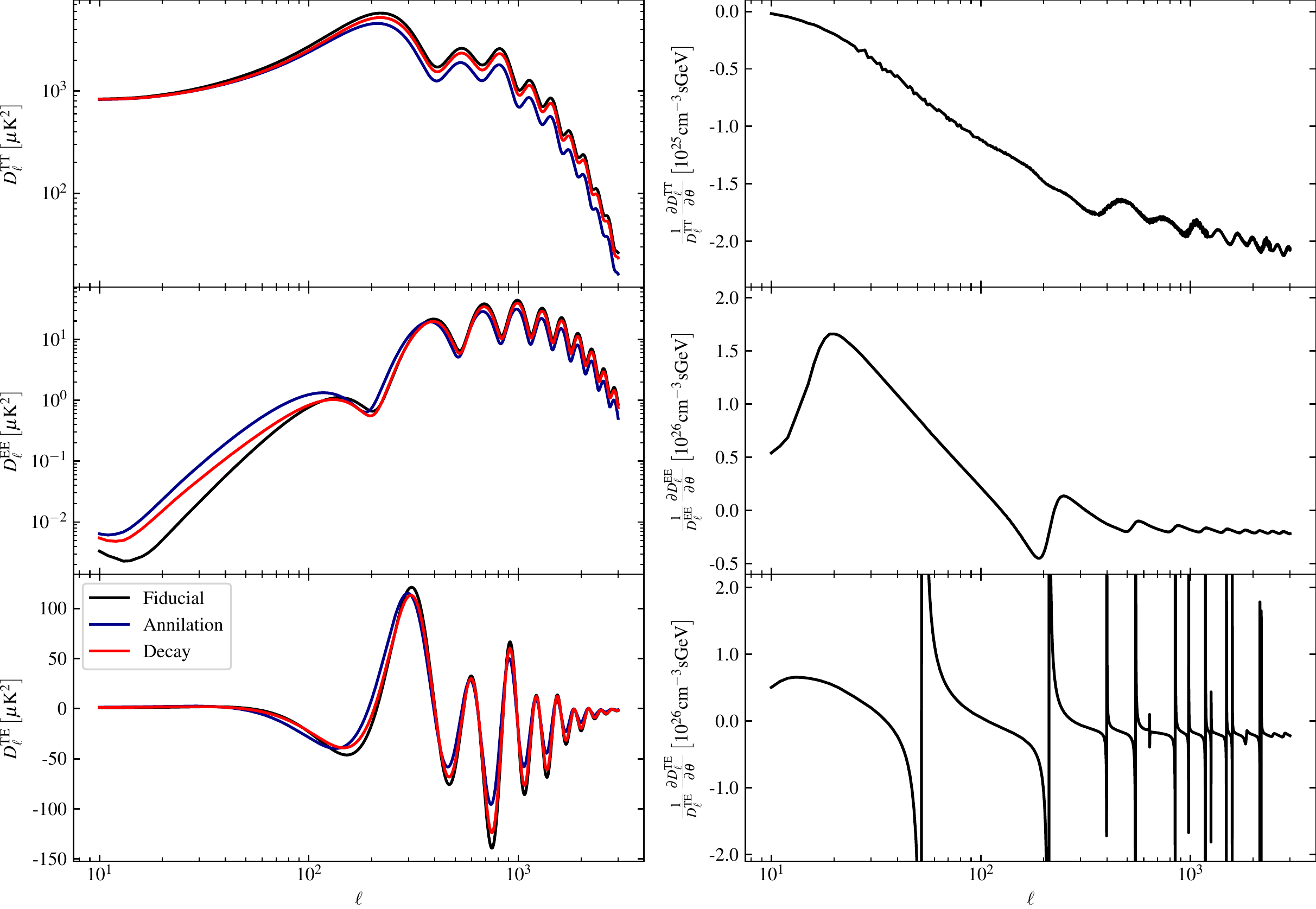}
    \caption{
    {\bf Left:} The impact of DM energy injection on CMB anisotropy power spectrum, 
    the top, middle and lower panels show $D^{\r{TT}}_{\ell}$, $D^{\r{EE}}_{\ell}$ and $D^{\r{TE}}_{\ell}$ respectively,
    the black solid lines show the $D_{\ell}$ in our fiducial $\Lambda \r{CDM}$ cosmology,
    whereas the red and blue lines show the $D_{\ell}$ for DM annihilating ($\pann = 10^{-26} \r{cm^3s^{-1}GeV^{-1}}$) and decaying ($\Gamma_{\chi} = 10^{-23}\r{s^{-1}}$) into $\gamma \gamma$,
    with the relevant DM mass $m_{\chi}$ set to 100 GeV.
    Legend applies to all panels.
    {\bf Right:}
    The spectrum normalized derivative of $D_{\ell}$ with respect to DM annihilation parameter $\theta \equiv \sv / m_{\chi}$ at fiducial value,
    assuming $m_{\chi} = 100 \r{GeV}$ and the anihilation channel considered is $\gamma \gamma$.
    }
    \label{fig:PS_variation}
\end{figure}

In presence of DM energy injection,
the evolution equations for ionisation level $\xe$ and gas temperature $\Tk$ takes the form~\cite{Liu:2016cnk,Slatyer:2016qyl}
\be
\frac{\rd \xe}
{\rd t}
=
\left[
\frac{\rd \xe}
{\rd t}
\right]_0 
+
\frac{1}{n_{\rm{H}}(z)E_{\r{i}}}
\left[\frac{{\rm{d}}E}{{\rm{d}}V{\rm{d}}t}\right]_{{\rm{dep,HIon}}}
+
\frac{1-C}{n_{\rm{H}}(z)E_\alpha}
\left[\frac{{\rm{d}}E}{{\rm{d}}V{\rm{d}}t}\right]_{{\rm{dep,Ly\alpha}}},
\label{dsfeffbshfdhvb76}
\ee
\be
\frac{\rd \Tk}
{\rd t}
=
\left[
\frac{\rd \Tk}
{\rd t}
\right]_0 
+ 
\frac{2}{3n_{\r{H}}(1+f_{\rm{He}}+x_{\rm{e}})}
\left[
\frac{{\rm{d}}E}
{{{\rm{d}}V}{\rm{d}}t}
\right]_{\rm{dep,Heat}},
\label{dsfeffbshfdhvb761}
\ee
here $\left[\rd \xe /\rd t \right]_0$ and $\left[\rd \Tk /\rd t \right]_0$ are the evolution equations in standard $\Lambda$CDM cosmology~\cite{Ali-Haimoud:2010hou,Seager:1999bc},
$n_{\r{H}}$ indicate hydrogen nuclei number density,
$E_{\r{i}} = 13.6 \r{eV}$ is the ionisation energy of neutral hydrogen,
$E_{\alpha} = 10.2 \r{eV}$ is the energy required to excite a ground stage hydrogen atom to the first excited state,
the factor $C$ describes the probability of an excited hydrogen to transit back to ground state~\cite{Liu:2016cnk,Slatyer:2016qyl,Seager:1999bc},
$f_{\r{He}}$ is the helium fraction by number of nuclei.

We modified the public {\tt HyRec} codes~\cite{Ali-Haimoud:2010hou} to account for DM contribution to ionisation and heating following Eqs. (\ref{dsfeffbshfdhvb76},\ref{dsfeffbshfdhvb761}),
and the corresponding CMB anisotropy spectra is computed using the {\tt CAMB} package~\cite{Lewis:1999bs}.
The heating from DM raises the gas temperature and can be constrained by 21cm observations~\cite{Lopez-Honorez:2016sur,Clark:2018ghm},
whereas the impact on CMB is mostly due to the ionisation effect,
which increases the number density of free electrons and thereby enhancing the Compton scattering between CMB photons and electrons.
As displayed in the left panels of Figure \ref{fig:PS_variation},
DM injection can lead to a supression on temperature anisotropy.
On the polarisation power spectrum,
DM can further enhance large scale anisotropy while shifting peak locations~\cite{Padmanabhan:2005es}.
The derivatives of the power spectrum with respect to the dark matter model parameters serves as a good indicator of CMB sensitivity to DM parameters, 
and we plot the normalised derivatives of CMB TT, TE and EE power spectrum with respect to DM annihilation parameter $\sv /m_\chi$ in the right panels of Figure \ref{fig:PS_variation}.

%% file: simulation_analysis.tex
%
In this section, 
we will start from the simulations of the microwave sky, 
including the CMB signals and foreground emission, 
we then simulate the response of two sets of configurations of microwave telescope to the sky signal,
and produce simulated observation maps. 
In contrast to the simulation and analysis process at the power spectrum level, 
we implement a component separation for the simulated sky maps to reconstruct the maps of the signal components, 
which is consistent with the analysis process for real observational data, 
and this analysis process provides the basis for a reasonable assessment of the foreground residual.
After obtaining the CMB angular power spectrum and its covariance, 
we use {\tt CosmoMC}~\cite{Lewis:2002ah,Lewis:2013hha} to estimate the sensitivities to the dark matter decay/annihilation parameters.

\subsection{Sky model simulation}\label{sec:datasimulation}

We have performed two simulations of the microwave sky using the model outlined in the Planck paper\cite{2016A&A...594A..10P}. 
In the following context, we will refer to these two foreground simulations as V1 and V2, respectively.
There is only a slight difference between these two simulations in terms of the models for synchrotron and thermal dust.
The signal on sky includes the CMB, the main galactic foreground emission with a diffuse foreground consisting of synchrotron radiation, free-free emission, thermal dust and spinning dust. 
Both temperature and polarized foreground maps were obtained using the Planck Sky Model (PSM) simulations \cite{2013A&A...553A..96D}.
Details of the sky model are given below:

\begin{enumerate}
    \item \textbf{CMB:} The input CMB maps are Gaussian realizations from a particular power spectrum obtained from the Boltzmann code CAMB~\cite{Lewis:1999bs}, using Planck 2018 best-fit cosmological parameters with $r = 0$. Throughout our simulation, we assume a spatially-flat $\Lambda \r{CDM}$ cosmology with the relevant parameters set by {\it{Planck}} 2018 results~\cite{Planck:2018vyg}:$h=0.6766$, $\Omega_{\r{b}} h^2=0.02242$, $\Omega_{\r{dm}} h^2=0.11933$, $\tau_{\r{reion}}=0.0561$, ${\r{ln}}(10^{10}A_{\r{s}})=3.047$, $n_{\r{s}}=0.9665$. The CMB dipole was not considered in this work.
    
    \item \textbf{Synchrotron:} Galactic synchrotron radiation arises from charged relativistic cosmic-rays accelerated by the Galactic magnetic field. For low frequency ($<$ 80 GHz), synchrotron is the dominant polarized foreground. The effects of Faraday rotation were ignored in our simulations, and we model the synchrotron emission across frequencies using the template maps according to specific emission laws. 
    For simulation V1, the frequency dependence of synchrotron emission is extracted from the \texttt{GALPROP} z10LMPD\_SUNfE synchrotron model\cite{2013MNRAS.436.2127O}. The model has no spatial dependence, meaning that each pixel on the sky has the same scaling factor.
    For another simulation V2, the synchrotron frequency dependence modeled as a power-law (in Rayleigh-Jeans unit):
    \begin{align}
        \begin{bmatrix}
            T_{sync}(\nu) \\ Q_{sync}(\nu) \\ U_{sync}(\nu)
        \end{bmatrix} \propto
        \begin{bmatrix}
            T_{template} \\ Q_{template} \\ U_{template}
        \end{bmatrix} \times \nu^{\beta_s}.
    \end{align}
    The typical value for spectral index $\beta_s$ is $-3$. For temperature maps, we use 408 MHz radio continuum all-sky map of Haslam et al.\cite{1982A&AS...47....1H, 2015MNRAS.451.4311R} as our template. And for polarization, the SMICA (Spectal Matching Independent Component Analysis) component separation of Planck PR3 (Planck Public Data Release 3)\cite{2020A&A...641A...4P} polarization maps are used as the templates with the same synchrotron spectral index.
    
    \item \textbf{Thermal dust:} For higher frequency ($>80$ GHz), the thermal emission from heated dust grains dominates the foreground signal. In PSM, the emission of thermal dust is modeled as the modified black-bodies\cite{2014A&A...571A..11P},
    \begin{align}
        I_{\nu}=A_\r{d}\nu^{\beta_\r{d}}B_{\nu}(T_\r{d}),
    \end{align}
    where $T_\r{d}$ is the temperature of the dust grains, $A_\r{d}$ is the amplitude and $\beta_\r{d}$ is the spectral index. $B_{\nu}(T_\r{d})$ is the Planck black-body function at temperature $T_\r{d}$ at frequency $\nu$. For polarization, the Stokes $Q$ and $U$ parameters (in ${\rm Mjy}/{\rm sr}$ unit) are correlate with the intensity, 
    \begin{align}
        Q_{\nu}(\hat{\mathbf{n}})&=f(\hat{\mathbf{n}}) I_{\nu}\cos(2\gamma_d(\hat{\mathbf{n}})), & 
        U_{\nu}(\hat{\mathbf{n}})&=f(\hat{\mathbf{n}}) I_{\nu}\sin(2\gamma_d(\hat{\mathbf{n}})),
    \end{align}
    where $f(\hat{\mathbf{n}})$ and $\gamma_d(\hat{\mathbf{n}})$ denote polarization fraction and polarization angle at different position $\hat{\mathbf{n}}$. 
    For V1 simulation, we use the GNILC (Generalized Needlet ILC) dust template at $353$ GHz in intensity and polarization of Planck PR3 as the template, which includes the template maps of the dust grains $T_\r{d}$, the amplitude for all $T,Q,U$ at 353 GHz, and the spectral index. In the V2 simulation, the same templates as those used in V1 are employed below 353~GHz. For frequencies above 353~GHz, the only distinction lies in the spectral index template. This template is derived from the modified blackbody fit of the dust spectral energy distribution (SED) for total intensity, obtained by applying the GNILC method to the PR2 HFI maps\cite{2016A&A...596A.109P}.
    
    \item \textbf{Free-free:} Free-free emission arises from electron–ion scattering in interstellar plasma. Generally, it is fainter than the synchrotron or thermal dust emission. Free-free is also modeled as power-law with a spectral index $\beta_\r{f}$ roughly $-2.12$. Free-free emission is intrinsically unpolarized. It contributes only to temperature measurements. The free-free emission map from \texttt{Commander} method \cite{2006ApJ...641..665E} in Planck PR2 \cite{2016A&A...594A..10P} are used as the template.
    
    \item \textbf{Spinning dust:} The small rotating dust grain can also produce microwave emission if they have a non-zero electric dipole moment. It can also be modeled as a power law within the CMB observation frequency bands. The template spinning dust map from \texttt{Commander} in Planck PR2 \cite{2016A&A...594A..10P} was used as the input template. Spinning dust emission is weakly polarized. In the simulation, the polarization fraction of spinning dust is set to be $0.005$.

    
\end{enumerate}

Overall, the V2 simulation is slightly more complex than V1, primarily due to its spatial dependence of the scale factor in synchrotron and the more intricate frequency dependence in thermal dust.
\subsection{Instrumental configuration}\label{sec:simulation_InstroConfig}

We consider two future CMB observing schemes: 
(1) A ground-based polarimetric telescope with a large array of superconducting transition-edge sensors (TES), similar to the Simons Small Aperture Polarimetric Telescope currently being deployed in the Southern Hemisphere\cite{2019JCAP...02..056A}, and the AliCPT experiment, currently under development and is to be deployed in Tibetan Ali of the Northern Hemisphere\cite{Li:2017drr,Li:2018rwc}. Both experiments use advanced ground-based small aperture polarimetric telescopes that carry out precise measurements of the CMB in the atmospheric frequency window. We also used simulated data from Planck HFI and WMAP K band, together with the ground test simulated data described above, for foreground removal analysis and subsequent parameter constraints. (2) A future space mission, the Probe of Inflation and Cosmic Origins (PICO), which is being designed to have even higher detection accuracy and to cover a wider frequency range\cite{2019arXiv190210541H,Aurlien:2022tlp}.
We use these two benchmark setups to examine the ability of near future CMB observations in detecting dark matter model parameters.

For simplicity,
we assume that the noise distribution of PICO is homogeneous among the patch without any specific scan strategy,
in this scenario the relevant instrumental noise power spectra $N_{\ell}$ is given by~\cite{Errard:2015cxa},
\be
N_{\ell}
=
\left[
\sum_{\nu}
N^{-1}_{\nu,\ell}
\right]^{-1}
,
\ 
N_{\nu,\ell} = \left(\frac{D_{\nu}}{b_{\nu,\ell}}\right)^2\frac{1}{1 \; \mathrm{sr} }.
\label{eq:noise_inst}
\ee
where $N_{\nu,\ell}$ is the noise power spectra at frequency $\nu$,
$D_\nu$ denotes the map-depth,
$b_{\nu,\ell}$ is the beam window function,
\begin{align}
 b_{\nu,\ell} = \exp\left(-\frac{1}{2}\frac{\ell(\ell+1)\theta^2_{\nu,{\rm FWHM}}}{8\ln 2}\right),   
\end{align}
and $\theta_{\nu,{\rm FWHM}}$ is the full width at half maximum (FWHM) beam size.
Table \ref{tab:setup} lists the experimental configurations of PICO used in this work.

\begin{figure}
\begin{center}
\subfloat[]{\includegraphics[width=0.32\textwidth]{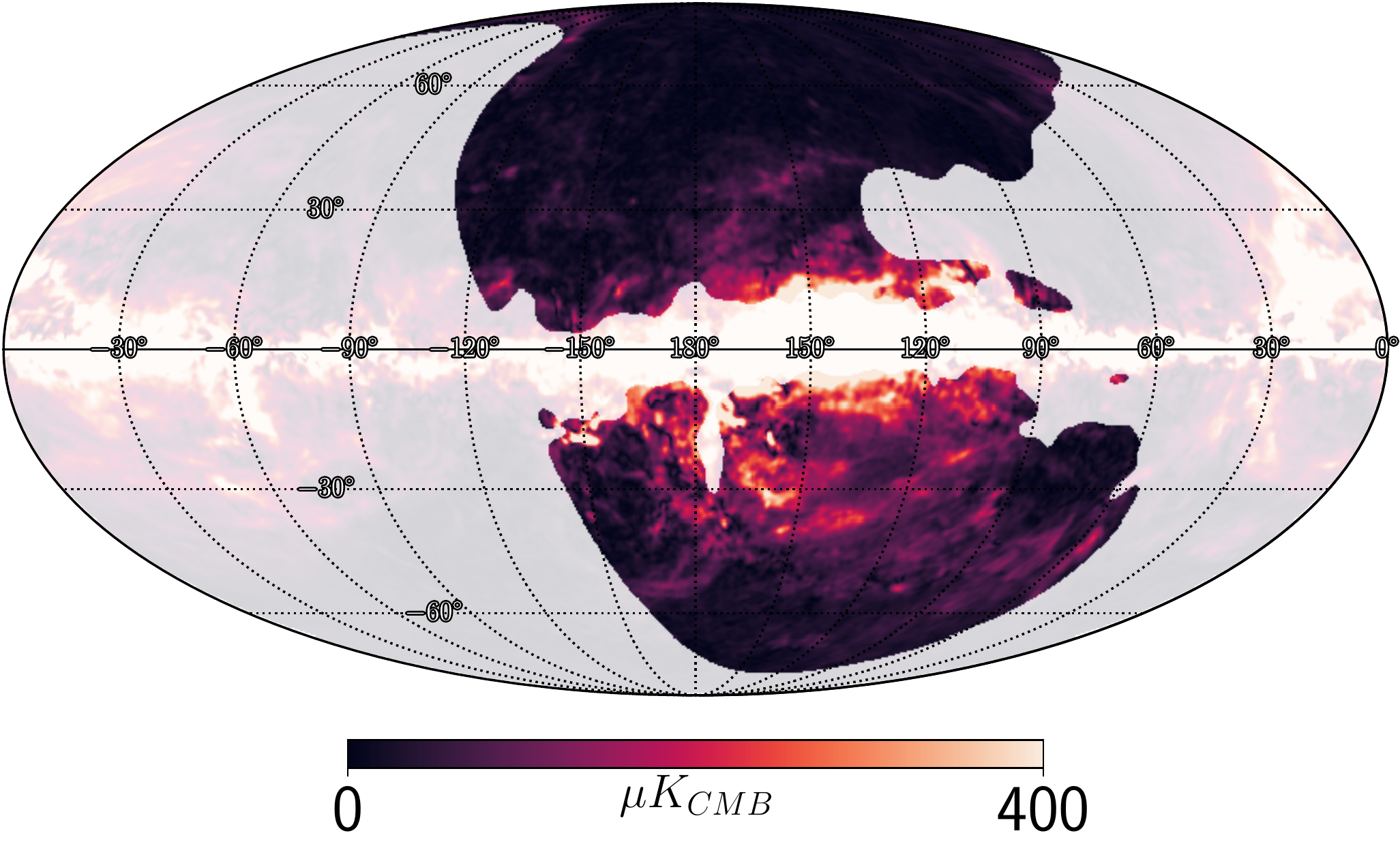}}\hskip 1mm
\subfloat[]{\includegraphics[width=0.32\textwidth]{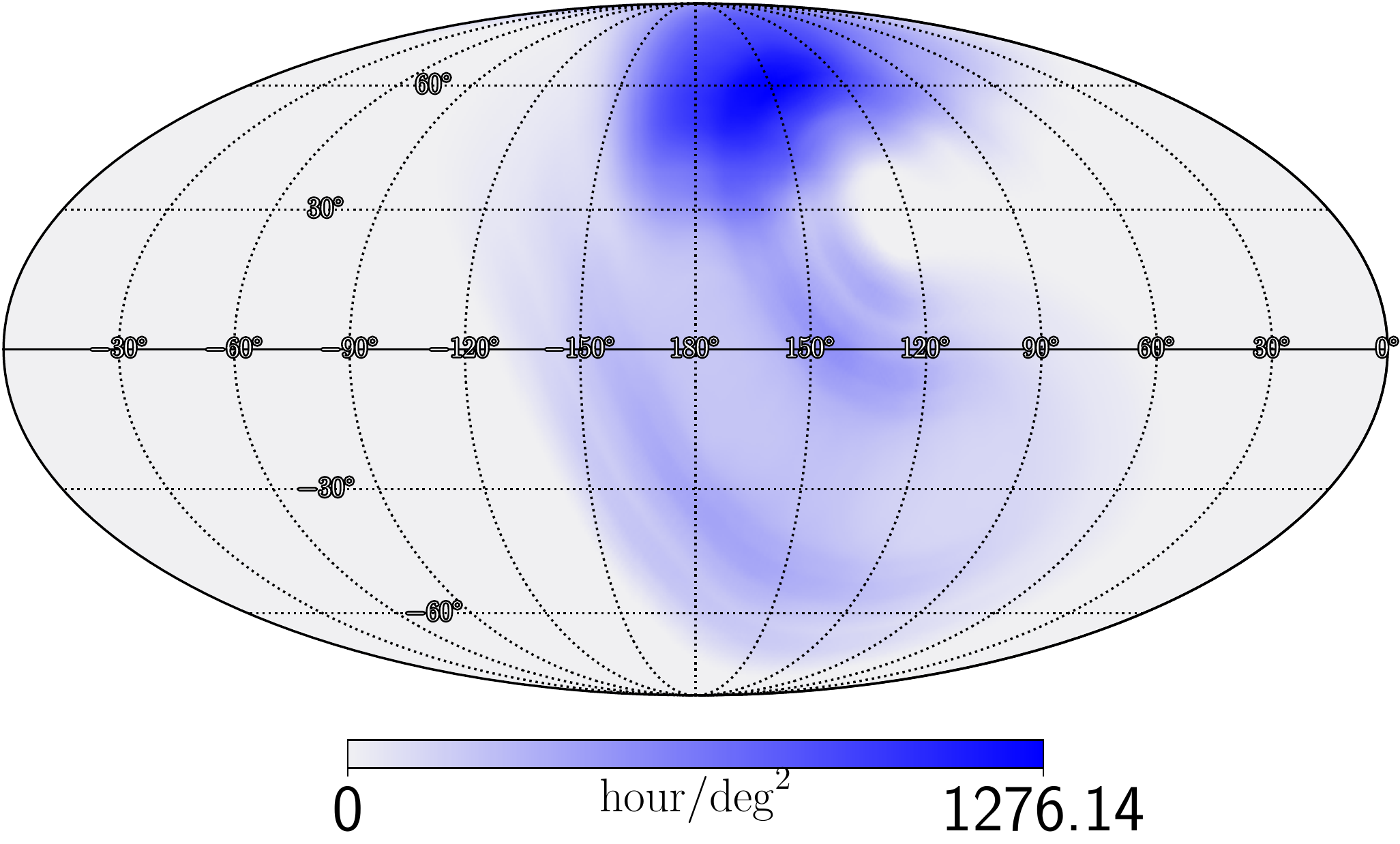}}\hskip 1mm
\subfloat[]{\includegraphics[width=0.32\textwidth]{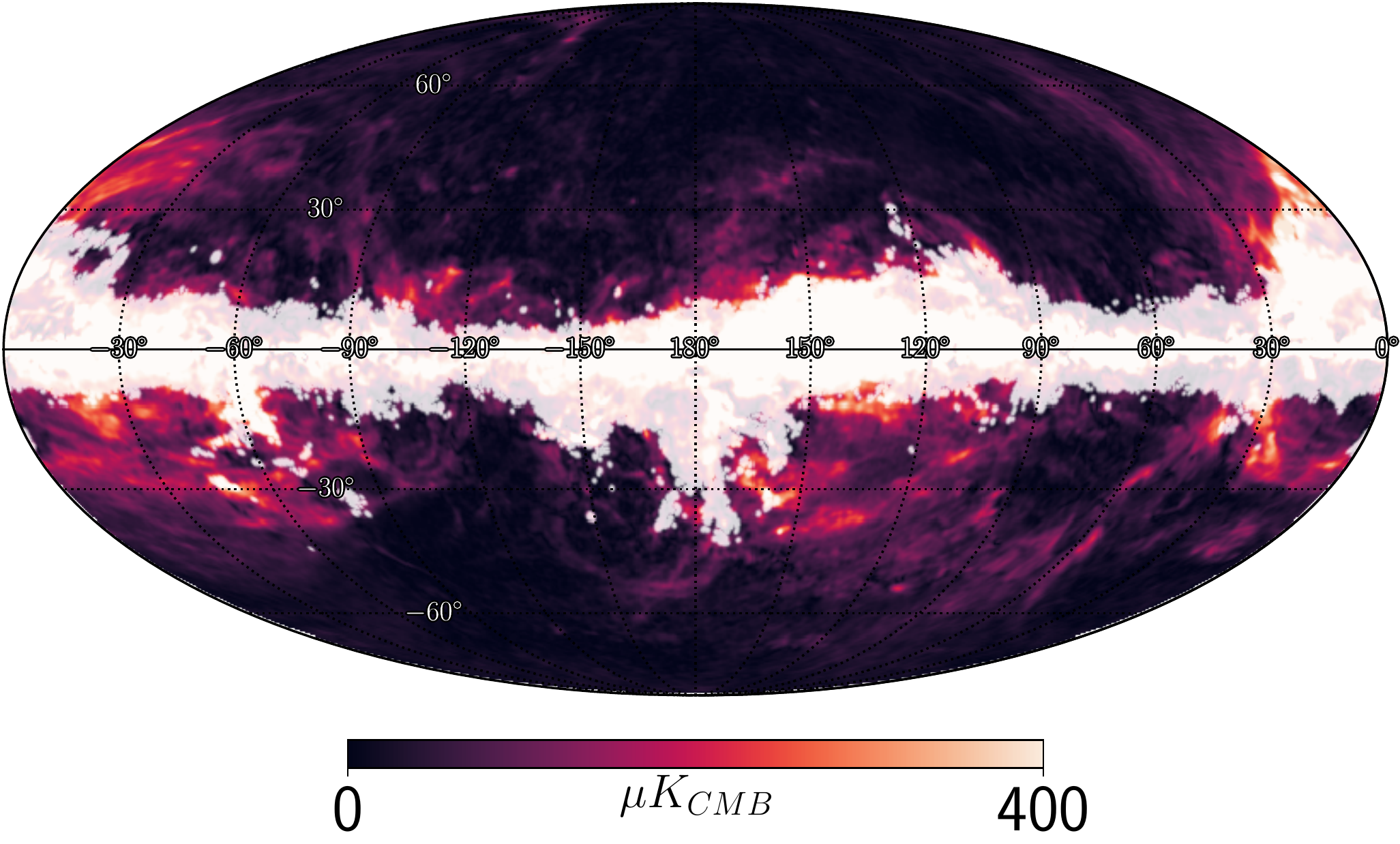}}
\end{center}
\caption{
(a): The sky patch used for the ground-based observations configuration. 
(b): Hit-map of the optimised observing strategy for the ground-based observations.
The hit-map is co-added by all detectors on focal plane. 
(c): The Galaxy mask used for PICO configuration. 
The background sky-map of sub-figure (a) and (c) is obtained from the Planck 353 GHz polarization emission intensity map $P = \sqrt{Q^2+U^2}$ \cite{2020A&A...643A..42P} with a resolution $1$~degree.
}
\label{fig:skypatch}
\end{figure}

For the ground-based observations, we adopt an optimised observing strategy covering a large sky area, which allows for as uniform a scan as possible. We compute the corresponding noise power spectra using the sky patches and hit-map in Figure \ref{fig:skypatch} and the noise equivalent temperature (NET) in Table \ref{tab:ground_noise_config}. It's easy to find out the variance of a pixel on noise simulation maps in terms of the hit-map and the NET,
\begin{align}
    \sigma^2_p = \frac{{\rm NET}^2}{H_p\Omega_p}.
\end{align}
$H_p$ denotes the value of hit-map at pixel $p$ (in ${\rm hour}/{\rm deg^2}$ unit) and $\Omega_p$ is the solid angle of pixel $p$. 
Compared with the PICO observation,
the noise of ground observation is inhomogeneous,
and the relationship between the noise power spectra $N_{\ell}$ and the map-depth is non-trivial,
thus equation \eqref{eq:noise_inst} is no longer applicable for this scenario.
In this work, 
we use $200$ noise-only simulation maps to estimate the $N_{\ell}$.
We generate the observed multi-frequency maps by adding the noise map realized from map depth and the simulated sky maps at each frequency. 
All the maps are pixelized in HealPix format at $\texttt{NSIDE}=1024$. 
We also use the same map depth to generate $100$ noise-only simulations which can be used to debias the noise at angular power spectral level.

\begin{table}[bthp]
    \caption{The instrumental specifications of PICO~\cite{2019arXiv190210541H}.Note that after subtracting the noisy Galactic plane, the effective sky coverage $f_{\r{sky}}$ for PICO is roughly 77\%.}
    \label{tab:setup}
    \centering
    \begin{threeparttable}
        \begin{tabular}{cccc}
            \toprule
            Freq.&Map depth T\tnote{*}&Map depth P& $\theta_{\r{FWHM}}$\\
            (GHz)&($\mu$K$\cdot$arcmin)&($\mu$K$\cdot$arcmin)&(arcmin)\\
            \midrule
            $21$&        $16.9$&    $23.9$&     $38.4$ \\
            $25$&        $13.0$&    $18.4$&     $32.0$ \\
            $30$&         $8.8$&    $12.4$&     $28.3$ \\
            $36$&         $5.6$&     $7.9$&     $23.6$ \\
            $43$&         $5.6$&     $7.9$&     $22.2$ \\
            $52$&         $4.0$&     $5.7$&     $18.4$ \\
            $62$&         $3.8$&     $5.4$&     $12.8$ \\
            $75$&         $3.0$&     $4.2$&     $10.7$ \\
            $90$&         $2.0$&     $2.8$&      $9.5$ \\
            $108$&         $1.6$&     $2.3$&      $7.9$ \\
            $129$&         $1.5$&     $2.1$&      $7.4$ \\
            $155$&         $1.3$&     $1.8$&      $6.2$ \\
            $186$&         $2.8$&     $4.0$&      $4.3$ \\
            $223$&         $3.2$&     $4.5$&      $3.6$ \\
            $268$&         $2.2$&     $3.1$&      $3.2$ \\
            $321$&         $3.0$&     $4.2$&      $2.6$ \\
            $385$&         $3.2$&     $4.5$&      $2.5$ \\
            $462$&         $6.4$&     $9.1$&      $2.1$ \\
            $555$&        $32.4$&    $45.8$&      $1.5$ \\
            $666$&       $125.2$&   $177.0$&      $1.3$ \\
            $799$&       $742.5$&  $1050.0$&      $1.1$ \\
            \bottomrule
        \end{tabular}
        \begin{tablenotes}
            \footnotesize
            \item[$*$] The map-depth of temperature map are computed from the map-depth of polarization map by dividing a $\sqrt{2}$ factor.
        \end{tablenotes}
    \end{threeparttable}
\end{table}

\begin{table}[bthp]
    \centering
    \caption{The instrumental configuration of the ground-based observation in this work, the map-depth, NET and beam-size are taken from the reference\cite{Li:2017drr} with $f_{\r{sky}}\simeq40\%$.}
    \label{tab:ground_noise_config}
    \begin{threeparttable}
        \begin{tabular}{ccc}
             \toprule
             Freq. & NET & Beam\\
             (GHz) & ($\mu {\rm K}\cdot \sqrt{s})$ & (arcmin)\\
             \midrule
             95 & 300 & 19 \\ 
             150 & 450 & 11 \\
             \bottomrule
        \end{tabular}

    \end{threeparttable}
\end{table}

\medskip

\subsection{Data processing for foreground removal}\label{sec:dataproc_ilc}

We adopt the Internal Linear Combination (ILC) methods for the component separation of the simulated maps.  
The CMB satisfies the Blackbody spectrum, which is not the case for the foreground emission, so that the CMB can be separated from foreground emission by means of measured data in different frequency bands. It is on the basis of this theoretical idea that the ILC achieves the separation of the components of the observed maps.
The ILC method is performed by implementing a weighted combination of observation maps for multiple channels, with the reconstructed maps satisfying a minimum variance, resulting in a set of weights. Since the CMB is frequency independent, the sum of the weighting coefficients of the CMB satisfies a value equal to unity. In this work, we employed two different ILC methods for the analysis, including ILC analysis in harmonic domain (HILC) and in Needlet domain (NILC). 

ILC method in harmonic domain removes the foreground by adding weights to each $\alm$. 
For a group of observed maps with $n$ frequency bands, 
each cleaned $\alm$ is defined by
    \begin{align}
        \alm = \sum_{i=1}^n w_\ell^i \alm^i
    \end{align}
Minimizing $\langle \alm^2 \rangle$ for every single $\ell$ with the constraint $\sum_{i} w_\ell^i = 1$ gives
    \begin{align}\label{ILC_weight}
        \mathbf{w}_{\ell} = \frac{\mathbf{C}_\ell^{-1} \mathbf{e}}{\mathbf{e^T} \mathbf{C}_\ell^{-1} \mathbf{e}}
    \end{align}
where $\mathbf{e}$ is a column vector of all ones,
$\mathbf{e^T}$ is the transpose of $\mathbf{e}$,
and $\mathbf{C}_\ell$ is the $n \times n$ matrix-valued cross-power spectrum

\begin{align}
    \mathbf{C}_{\ell}^{ij} = \langle a_{\ell m}^i a_{\ell m}^{j*} \rangle
\end{align}

ILC method in needlet means implement the ILC on a frame of spherical wavelets, (needlets). The characteristic of this method is that it can take localization in spatial domain and harmonic domain simultaneously. Decompose the maps on each channel c $a_{\ell m}^c$ into a set of filtered maps $a_{\ell m}^{c,j}$ 
\begin{align}
    a_{\ell m}^{c, j}=h_l^j a_{\ell m}^c
\end{align}
the spherical needlets are defined as
\begin{align}
    \Psi_{j k}(\hat{n})=\sqrt{\lambda_{j k}} \sum_{\ell=0}^{\ell_{\max }} \sum_{m=-\ell}^{\ell} h_{\ell}^j Y_{\ell m}^*(\hat{n}) Y_{\ell m}\left(\hat{\xi}_{j k}\right)
\end{align}
where the $h_{\ell}^j$ is the filters which can keep the harmonic information, which satisfies $\sum_j\left(h_{\ell}^j\right)^2=1$ , $\sqrt{\lambda_{j k}}$ are weights of every grid points, and$\left(\hat{\xi}_{j k}\right)$ a set of grid points on the sphere for scale j. The needlet coefficients for a CMB field $X(\hat{n})$ are denoted as 
\begin{align}
    \beta_{j k}=\int X(\hat{n}) \Psi_{j k}(\hat{n}) \mathrm{d} \Omega_{\hat{n}}
\end{align}
Similarly, linearly combine maps on different frequency bands, the needlet coefficients of cleaned CMB map is
\begin{align}
    \hat{\beta}_{j k}=\sum_{c=1}^{n_\nu} w_{j k}^c \beta_{j k}^c
\end{align}
where $n_\nu$ is the number of channels, $w_{j k}^c$ is the needlet weight for scale j and frequency channel c, at the pixel k of the HEALPix representation of the needlet coefficients for that scale, satisfies
\begin{align}
    \sum_{c=1}^{n_{c}} a^{c} \omega_{j k}^{c}=1
\end{align}
The needlet ILC weights can finally be expressed as
\begin{align}
    \widehat{\omega}_{j k}^c=\frac{\sum_{c^{\prime}}\left[\widehat{R}_{j k}^{-1}\right]^{c c^{\prime}} a^{c^{\prime}}}{\sum_c \sum_{c^{\prime}} a^c\left[\widehat{R}_{j k}^{-1}\right]^{c c^{\prime}} a^{c^{\prime}}}
\end{align}
where ${R}_{j k}^{-1}$ is an estimate of the inverse covariance.
As we want to remove foreground or noise at different localized scale the covariance matrix are an estimate as an average of the product of the relevant computed needlets coefficients over some local space
\begin{align}
    {R}_{j k}^{c c^{\prime}}=\frac{1}{n_{k}} \sum_{k^{\prime}} w_{j}\left(k, k^{\prime}\right) \beta_{j k}^{c} \beta_{j k}^{c^{\prime}}
\end{align}
Finally,  the NILC CMB map is
\begin{align}
    X^{\mathrm{NILC}}(\hat{n})= & \sum_{\ell m} a_{\ell m}^{\mathrm{NILC}} Y_{\ell m}(\hat{n}) \notag \\
    = & \sum_{\ell m} \left( \sum_{j} \sum_{k}  \sqrt{\lambda_{j k}} \hat{\beta}_{j k}h_{\ell}^{j} Y_{\ell m}\left(\hat{\xi}_{j k}\right) \right) Y_{\ell m}\left(\hat{n}\right)
\end{align}


We begin our implementation first by enforcing the maps at all frequencies to share the same beam size,
then deconvolve their own beam sizes. 
For the simulated ground-based observation, we perform foreground removal only once on the data using the ILC method. 
For the PICO mission, the full sky coverage maps, 
we implemented a more refined foreground cleaning analysis.
Following the considerations in Ref.~\cite{Tegmark:2003ve} for WMAP data processing,
we take into account the foreground variability in different sky regions in order to obtain better foreground removal.
Specifically,
we divided the map into three different regions according to their brightness of the foreground emission and implemented ILCs separately.
To avoid mirages of foreground emission from the Galactic plane leaking up to high latitudes, we start the component separation from regions with higher foreground levels, and then move on to the cleaner areas. We first clean the dirtier area individually and replace the corresponding region of the temporary maps by the cleaned map, and then clean the new sky map after replacement.

\subsection{Data processing for constraining dark matter parameters}
\label{sec:dataproc_constraints}

After foreground removal,
the simulated data is composed of CMB,
the foreground residual and instrumental noise, and the cleaned spherical harmonic coefficient $\alm$ is given by
\begin{align}
\alm = a_{\ell m,{\rm CMB}} + a_{\ell m,\r{res.}} + a_{\ell m,\r{noise}}.
\label{eq:hilc}
\end{align}
For HILC method, the foreground residual and instrumental noise residual can be expressed as
\be
a_{\ell m,\r{res.}}^{\mathrm{HILC}} = \sum_{i=1}^n w_\ell^i a_{\ell m,{\rm res.}}^i,\ 
a_{\ell m,\r{noise}}^{\mathrm{HILC}} = \sum_{i=1}^n w_\ell^i a_{\ell m,{\rm noise}}^i,
\label{ds8uei3gsfe5}
\ee
and for2 NILC method
\begin{align}
    a_{\ell m, \r{res.}}^{\mathrm{NILC}}=\sum_j \sum_k \sqrt{\lambda_{j k}} \beta_{j k, \r{res.}} h_l^j Y_{l m}\left(\hat{\xi}_{j k}\right)
\end{align}
\begin{align}
    a_{\ell m, \r{noise}}^{\mathrm{NILC}}=\sum_j \sum_k \sqrt{\lambda_{j k}} \beta_{j k, \r{noise}} h_l^j Y_{l m}\left(\hat{\xi}_{j k}\right)
\label{nilcfgr}
\end{align}
Since the cross correlation between CMB, foreground residual and the instrument noise vanishes,
the observed power spectra can be written as
\be
C^{\r{xy}}_\ell
=
C^{\r{xy}}_{\ell,\r{CMB}}
+
C^{\r{xy}}_{\ell,\r{res.}}
+
N^{\r{xy}}_{\ell,\r{noise}},
\label{douywiydsf4}
\ee
\be
C^{\r{xy}}_{\ell,\r{CMB}}
=
\frac{1}{2 \ell +1}
\sum_{m = -\ell}^{\ell}
\left<a^{\r{x}}_{\ell m, \r{CMB}} \cdot a^{\r{y*}}_{\ell m, \r{CMB}}\right>,
\ee
\be
C^{\r{xy}}_{\ell,\r{res.}}
=
\frac{1}{2 \ell +1}
\sum_{m = -\ell}^{\ell}
\left<a^{\r{x}}_{\ell m, \r{res.}} \cdot a^{\r{y*}}_{\ell m, \r{res.}}\right>,
\ee
In this work we only consider temperature (T-mode) and E-mode polarisation correlations,
thus here $\r{xy} \in [\r{TT, TE, EE}]$.
For the PICO mission,
$N^{\r{xy}}_{\ell,\r{noise}}$ in \Eq{douywiydsf4} is computed following \Eq{eq:noise_inst},
whereas for the ground observations,
$N^{\r{xy}}_{\ell,\r{noise}}$ is calculated by
\be
N^{\r{xy}}_{\ell,\r{noise}}
=
\frac{1}{2 \ell +1}
\sum_{m = -\ell}^{\ell}
\left<a^{\r{x}}_{\ell m, \r{noise}} \cdot a^{\r{y*}}_{\ell m, \r{noise}}\right>.
\ee
Note that the noise TE cross-correlation $N^{\r{TE}}_{\ell,\r{noise}}$ vanishes for both PICO and ground observations.

To derive constraints on DM,
we constructed the mock likelihood as~\cite{Hamimeche:2008ai},
\be
-2{\ln}\mathcal{L} 
= f_{\rm sky} \sum_{\ell}(2\ell +1)\left[{\rm{Tr}}({\hat{C}_{\ell}}C_{\ell}^{-1}) -\ln|{\hat{C}_{\ell}}C_{\ell}^{-1}|-2\right],
\label{Likeldsfeyihood}
\ee

\be
C_{\ell} \equiv 
\begin{bmatrix}
C_{\ell}^{\rm TT} & C_{\ell}^{\rm TE} \\
C_{\ell}^{\rm TE} & C_{\ell}^{\rm EE}\\
\end{bmatrix},
\ 
\hat{C}_{\ell} \equiv 
\begin{bmatrix}
\hat{C}_{\ell}^{\rm TT} & \hat{C}_{\ell}^{\rm TE} \\
\hat{C}_{\ell}^{\rm TE} & \hat{C}_{\ell}^{\rm EE}\\
\end{bmatrix},
\label{Signal_CLsaduiy}
\ee
here $C_{\ell}^{\rm xy}$ and $\hat{C}_{\ell}^{\rm xy}$ are theoretical and fiducial CMB power spectra plus foreground residual and noise respectively (see \Eq{douywiydsf4}).
For ground observation and PICO,
we consider multipole ranges of $\ell \in [30,620]$ and $\ell \in [10,3000]$ respectively.
The CMB signal in both the fiducial and theoretical spectra are calculated using the {\tt CAMB}~\cite{Lewis:1999bs} package,
which has been interfaced with our modified {\tt HyRec}~\cite{Ali-Haimoud:2010hou} to account for the effects of DM energy injection.
Our fiducial CMB power spectra is computed for a flat $\Lambda \r{CDM}$ cosmology detailed in Section ~\ref{sec:datasimulation}.
Unless specified otherwise,
our DM constraints are set by performing MCMC (Monte Carlo Markhov Chains) analysis on the likelihood in \Eq{Likeldsfeyihood} using {\tt CosmoMC}~\cite{Lewis:2002ah,Lewis:2013hha} codes.
In addition to the DM parameters,
the base $\Lambda \r{CDM}$ parameters are also varied during MCMC and we marginalise them when presenting our DM limits.
As a consistency check,
we also compared the results of {\tt CosmoMC} with constraints set by {\tt EMCEE} sampler~\cite{Foreman-Mackey:2012any} and a simple Fisher matrix formalism,
and we found that all these approaches give consistent results.

%% file: results.tex
%
Foreground residuals will be present after component separation and mix into CMB maps. These residuals contribute to extra uncertainties in the CMB spectrum, ultimately biasing the physical parameter fits. 
In this section, we start with the results of component separation and foreground subtraction of analogue maps, and discuss the impact of foreground residuals on the parameter uncertainties on the fitting of dark matter annihilation/decay models.


\subsection{Foreground residual}\label{Fg_effects}

\begin{figure}[!h]
\begin{center}
{\includegraphics[width=\textwidth]{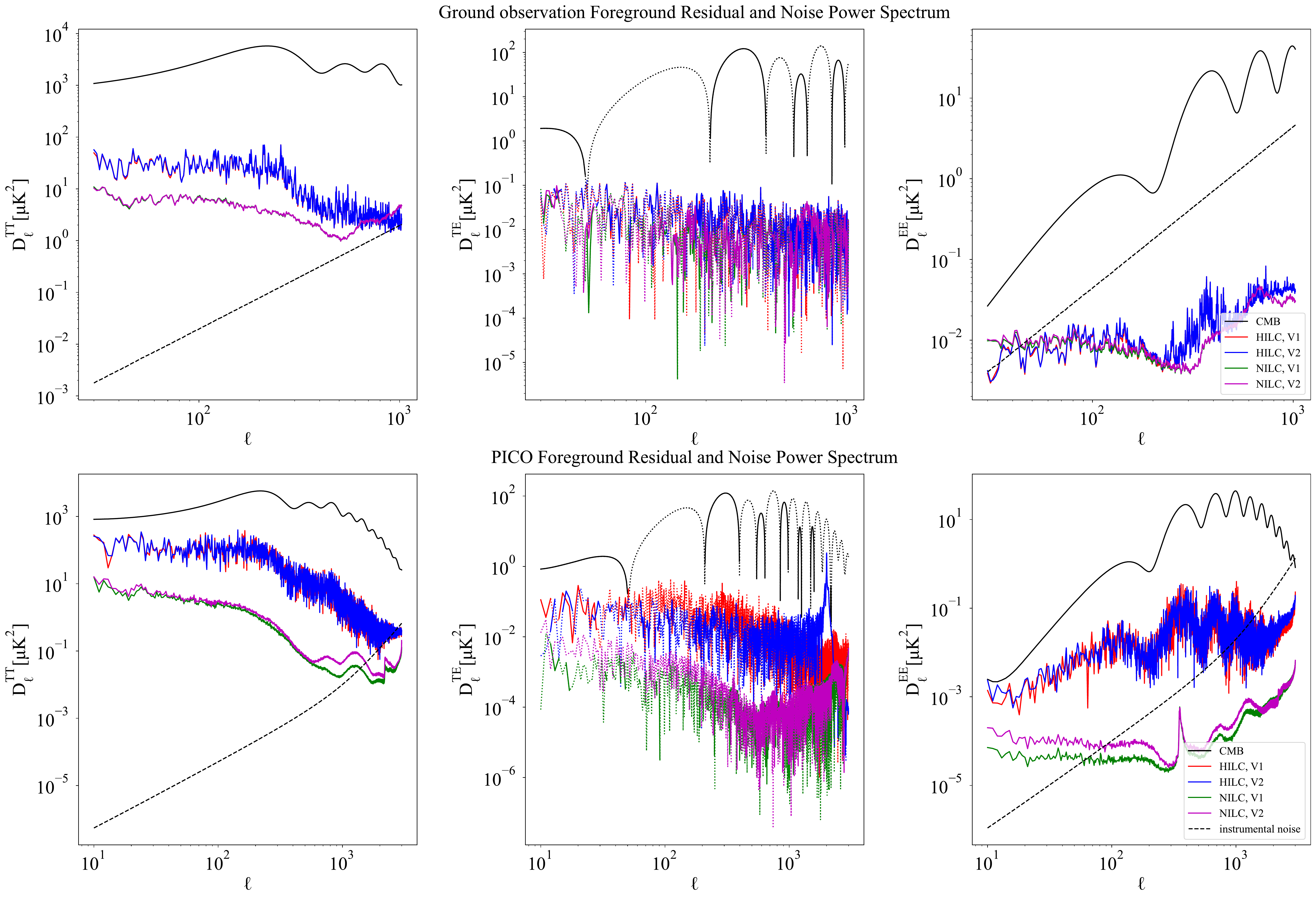}}        
\end{center}    
\caption{
Comparison of CMB power spectrum (black solid), 
foreground residual (coloured solid) and instrumental noise (dashed).
The left, middle and right panels correspond to TT, TE and EE correlations respectively. The top and bottom panels correspond to the ground observation and PICO respectively. 
In the middle panels,
we indicate the positive correlation with solid curves,
whereas negative correlation is shown in dotted lines.
Legend applies to all panels.
}
\label{fig:fgResiduals}
\end{figure}

As stated in the equations \eqref{ds8uei3gsfe5} and \eqref{nilcfgr}, the foreground residual is proportional to the foreground emission term. Thus intense emission leads to a high residual after ILC. 
The upper and lower panels in Figure \ref{fig:fgResiduals} show the foreground residual power spectra of TT, TE and EE after performing the ILC analysis of the two different missions respectively, and the CMB power spectrum are also included in black solid lines for comparison.
The two missions are introduced in Section~\ref{sec:simulation_InstroConfig}, 
of which one for the ground-based telescope which realizes partial sky observation in northern hemisphere (red), 
and the other one for the future PICO's multi-band and full-sky observations (blue). 
The noise power spectra for the two configurations are also shown in Figure \ref{fig:fgResiduals} as dashed lines for comparison.

It is clear that in the TT spectra the foreground residual are much larger than the level of the instrumental noises, 
especially for $\ell<1000$, 
where the foreground residuals are two to three orders of magnitude higher. 
Thanks to its low-noise and multi-band configuration,
PICO exhibits lower foreground residual than the ground observation,
especially at small scales.
For the TE cross spectrum, 
the instrumental noise vanishes and thus foreground residual is the only contamination term to the observed signal.
For EE, 
while the foreground residual is higher on large scales than the instrumental noise power spectra,
yet interestingly, 
on small scales it is subdominant to noise. 
Judging from the relative height between the foreground and noise levels in TT and EE, 
one would expect the foreground residuals to be a major source of uncertainty if physics fitting heavily relies on the TT spectrum, 
and much less in case EE data dominate the fitting sensitivity. 

Comparing the results obtained by two different foreground removal methods, we can see that the foreground residual obtained by applying the NILC method is lower than that of the HILC method, especially on PICO with more frequency bands and lower instrumental noise level, the foreground residual level given by NILC is one to two orders of magnitude lower than that of HILC. For the ground observation, the difference between the two is slightly smaller, about an order of magnitude. In general, due to the characteristics of the NILC method, which is localized in the pixel space and the spherical harmonic space at the same time, it has better performance in foreground removal than HILC, and will get a lower foreground residual.


Using different foreground removal methods, working on different foreground models, we get some interesting results by comparison.
As mentioned in Section~\ref{sec:datasimulation}, V2 is more complex than V1, as comparing with V1, V2 considers the spatial dependence of synchrotron, and it adopts a different spectral index template in the frequency band above 353~GHz for the simulation of thermal dust. 
For the ground observation simulation data, since the frequency above 353~GHz is not considered, the difference between the two foreground models is only synchrotron radiation, which is the dominant item in the foreground in the frequency below 80~GHz, and most of the ground observation frequency bands we consider are above 80~GHz , the foreground is mainly dominated by dust that is consistent in the two models, so the calculated foreground residual results are similar.
For PICO, as can be seen from the results,  the foreground residual obtained by performing NILC on V2 is slightly higher than that of V1, but if use the HILC method to remove the foreground, the difference in the results is very small. The reason is that V2 mainly increases the space complexity compared with V1, and this has little effect on the HILC performed in spherical harmonic space, while the result of using NILC that considers both pixel space and spherical harmonic space will be produce visible changes.

\begin{figure}[h]
\begin{center}\label{fig:diffConstraits}
{\includegraphics[width=\textwidth]{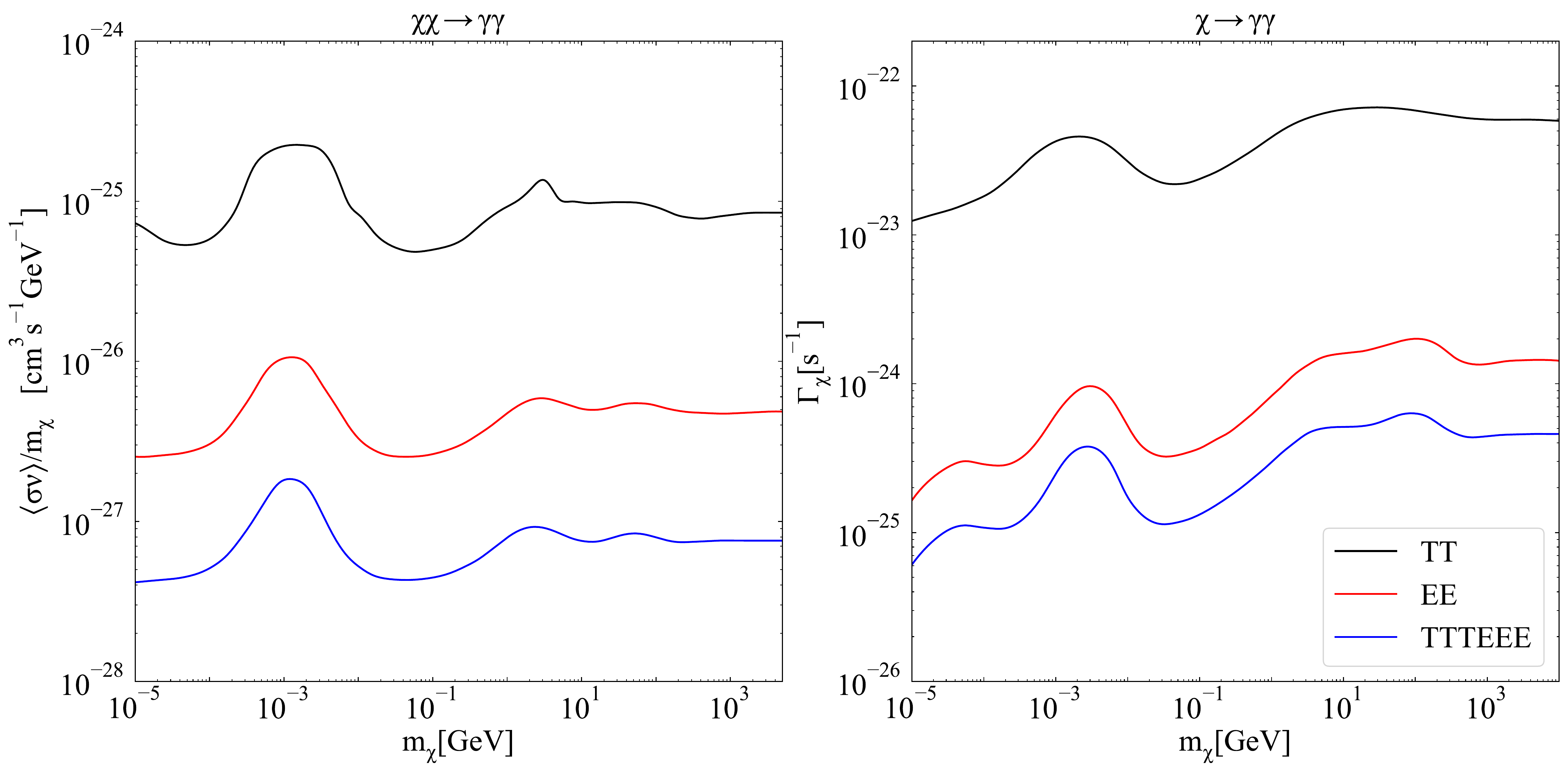}}
\end{center}  
\caption{
Marginalised 95\% C.L. upper bounds on DM annihilation parameter $\sv / m_\chi$ (left) and decay width $\Gamma_\chi$ (right),
set by simulated TT (black), EE (red) and TT+TE+EE (blue) datasets for ground observations, using foreground V1 and NILC to remove foreground.
DM is assumed to annihilate/decay into $\gamma \gamma$, legend applies to both panels.
}\label{fig:diffConstraits}
\end{figure}

To illustrate the impact of the foreground residual on DM constraints, 
we first analyse the independent limits from temperature and polarisation respectively.
Figure \ref{fig:diffConstraits} shows ground observation constraints on DM annihilating/decaying into $\gamma\gamma$, using NILC method and foreground model V1.
We found that the limits set by the EE spectrum are an order of magnitude more stringent than that set by the TT spectrum alone,
and combining TT, EE and TE further improves the constraint by a factor of two compared to the EE spectrum.

This result is understandable from DM imprint's characteristics in different spectra. 
The main effect of extra energy release is to raise the ionization fraction which broadens the last scattering surface, 
leading to a suppression of the anisotropy amplitude in both temperature and polarisation maps. 
This damping effect has degeneracy with a few $\Lambda$CDM parameters, esp. the optical depth $\tau_{\rm reion}$. 
Since the cosmic reionization history has not been accurately measured, 
the large uncertainty in the modeling of astrophysical ionization ($z<20$) process prevents TT data alone from giving a stringent constraint. 
Currently, 
a meaningful sensitivity of DM energy injection mostly derives from EE maps, 
in which the $\ell-$location shift of the leading resonance peaks are more visible than that in TT maps~\cite{Padmanabhan:2005es}.
This shift can not be mimicked by damping attenuation or simple deformation in the primordial spectrum, 
and makes the polarisation power spectrum more helpful in breaking the correlation between DM annihilation and that from the underlying $\Lambda$CDM parameters, 
such as $n_\r{s}$ and $A_\r{s}$~\cite{Padmanabhan:2005es}. 
In addition, 
on a very wide angular scales, 
where $\ell$ is between $100$ and $2000$, 
dark matter produces an enhancement of polarization due to an increase in the ionization fraction of the frozen tail after recombination. 

\begin{figure}[t]
\begin{center}
{\includegraphics[width=\textwidth]{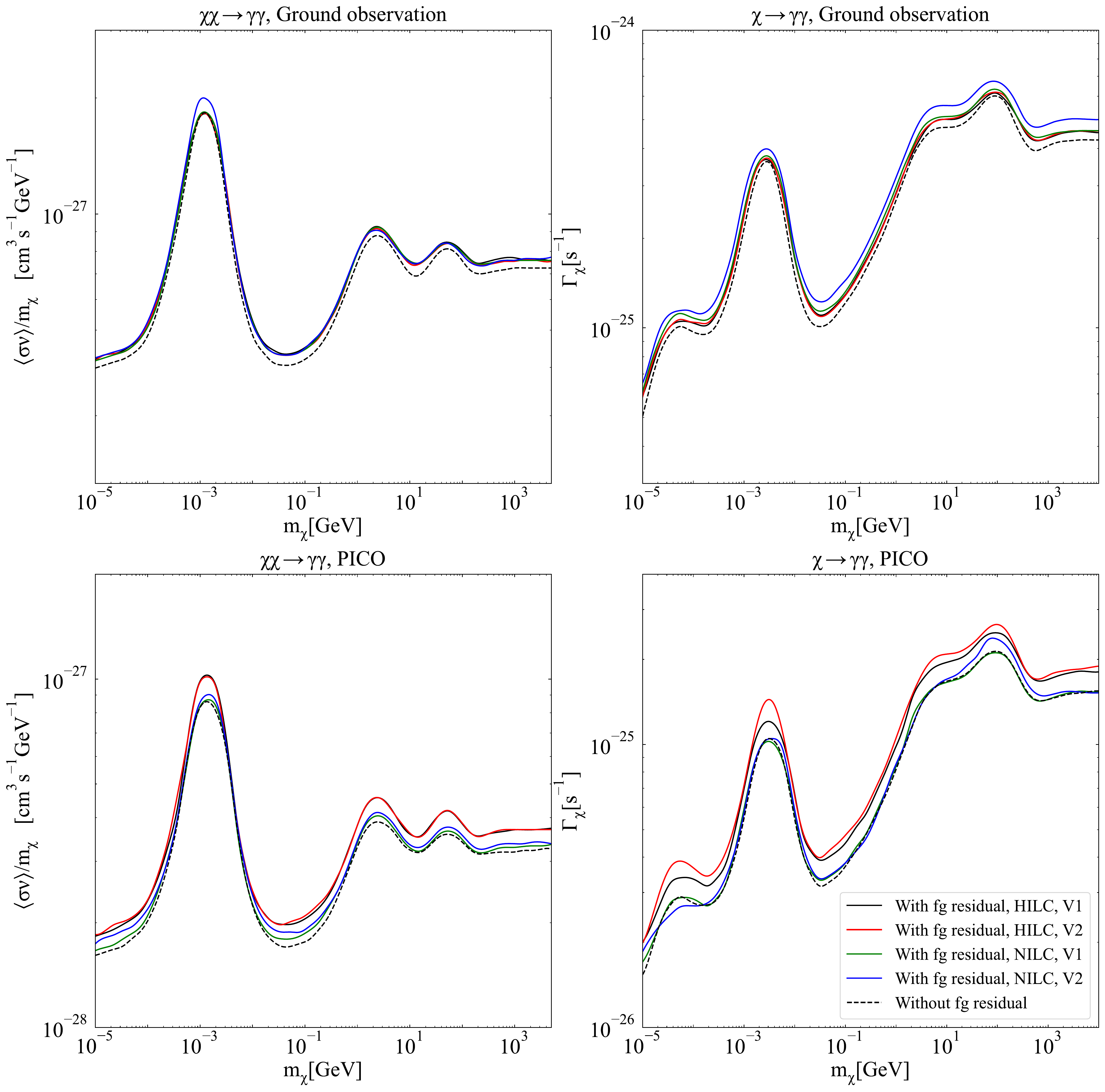}}        
\end{center}
\caption{
The impact of foreground residual on DM parameter constraints.
The solid and dashed curves show results with and without foreground residual respectively,
regions above the lines are excluded at 95\% C.L.. The upper and lower panels shows limits set by ground observation and PICO simulated data respectively.
The left panel shows constraints on annihilation parameter $\sv /m_\chi$,
and the right panel represents limits on decay width $\Gamma_\chi$.  
DM annihilation/decay channel considered here is $\gamma \gamma$,
the legend applies to both panels.
Compared to noise-only limits, depending on the foreground model and foreground removal method, 
presence of foreground residual weakens the sensitivity on annihilation by an average of 9\% and 6\% - 16\% for ground-base and PICO configurations. 
In the decay channel, 
for the ground observation, the foreground residual's average sensitivity reduction is 7\%-19\%, and for PICO, results can be increased by up to 12\%.
}
\label{fidsuyysfggff}
\end{figure}

Figure \ref{fidsuyysfggff} illustrates the impact of foreground residual on DM constraints,
with the solid and dashed lines representing the results with and without foreground residual respectively.
The foreground residual on the TT spectrum introduces minor changes in parameter constraints because the constraints from the TT spectrum are not dominant terms, 
even though the foreground residuals are quite large compared to instrumental noise. 
However, 
the case is different for the EE correlation, 
where the strength of the foreground residual is comparable to the noise level of the instrument.
Since EE plays a dominant role in the parameter constraint, 
the foreground residual in the EE spectrum bring about a more significant effect. 
For the ground observations,
foreground residual can weaken the DM constraints by about 7\%-19\%,
whereas for the PICO mission,
the impact of foreground residual is less significant, for the highest foreground residual level provided by HILC method on V2, 16\%(annihilation) and 12\%(decay)of the constraints were weaken as a result, and for the case of using NILC method and foreground V1, the results are almost consistent with the case with no foreground residual.
Mainly thanks to its multi-frequency observation strategy and the lower instrumental noise of PICO, 
which makes the separation of components on map level more efficient, 
 therefore the foreground residual of the polarisation spectrum is not dominant on some scales. 

\begin{figure}
\begin{center}\label{fig:results}
{\includegraphics[width=\textwidth]{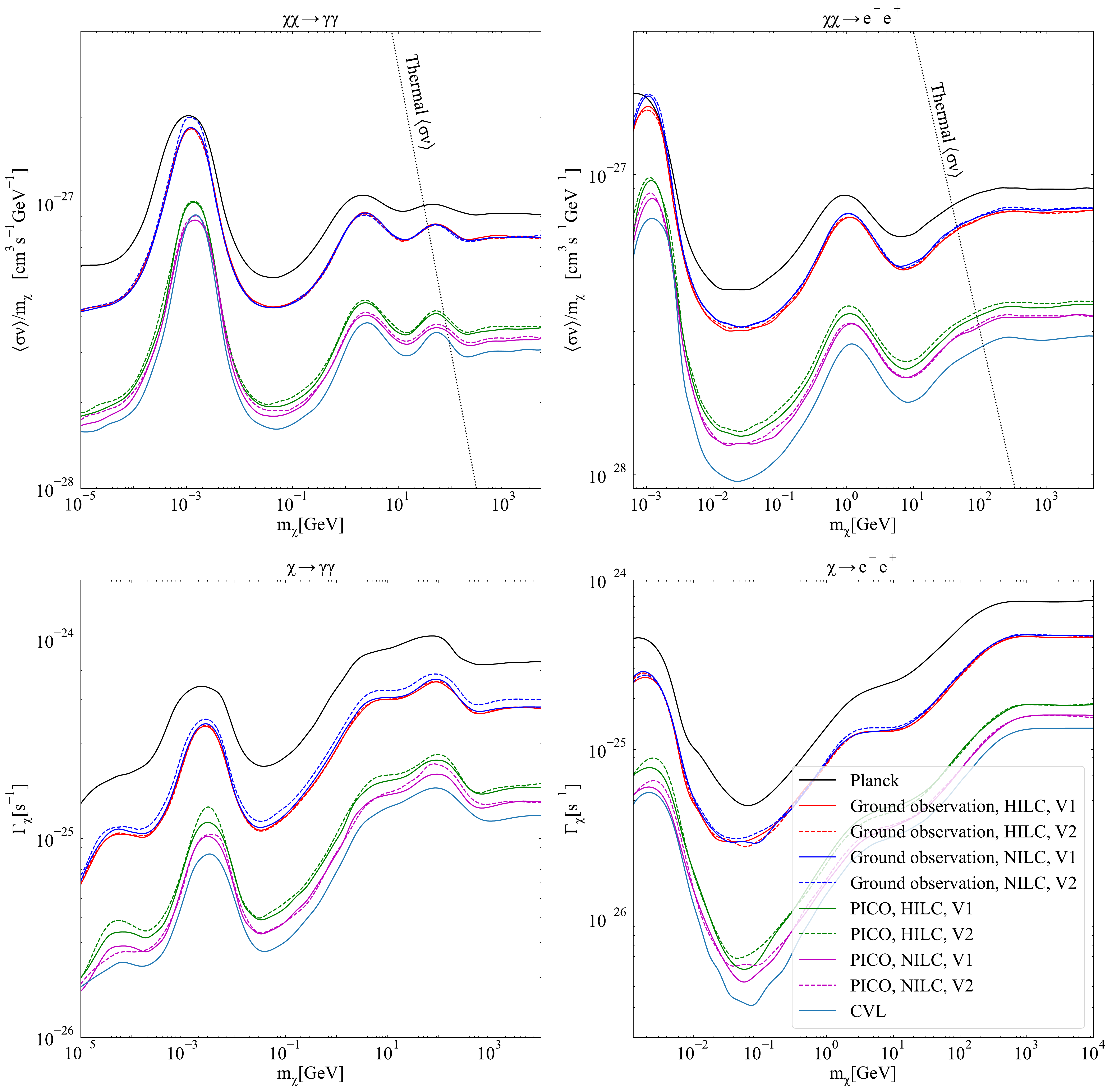}}
\end{center}
\caption{
95\% C.L. upper bounds on DM annihilation parameter $\sv /m_\chi$ (top) and decay width $\Gamma_\chi$ (bottom).
The left and right panels correspond to DM annihilation/decay channels of $\gamma \gamma$ and $\r{e^{\pm}}$ respectively.
The black lines show limits set by current {\it Planck} 2018 TT+TE+EE likelihoods, and the cyan curves show the constraints from a cosmic variance limited (CVL) experiment observing in multipole range of $\ell \in [10,3000]$.
The dotted lines on the top panels show the thermal relic cross-section of $\sv = 3 \times 10^{-26} \r{cm^3 s^{-1}}$. Legend applies to all panels.
}
\label{fig:pConstraints}
\end{figure}

\subsection{Constraints on dark matter parameters}\label{Para_constraits}

In this section we show the prospective limits on the dark matter annihilation and decay rate parameters from the simulated data. 
Both $\sv /m_\chi$ and $\Gamma_\chi$ are time and velocity independent quantities in our calculations, 
and our constraints are obtained by sampling the mock likelihood in \Eq{Likeldsfeyihood} using {\tt CosmoMC}~\cite{Lewis:2002ah,Lewis:2013hha}.
In Figure \ref{fig:pConstraints}, 
we show the results of the calculations, 
where the uncertainties in the power spectrum include both instrumental noise and foreground residuals. 
The upper and lower panels show the 95\% C.L. upper bounds on $\sv /m_{\chi}$ and $\Gamma_\chi$ respectively,
and the left and right panels show constraints for $\gamma \gamma$ and $\r{e^{\pm}}$ channels respectively.
We compare the limits given by TT, TE and EE measurements from four different experiments,
the black lines shows the results set by Planck 2018 likelihoods~\cite{Planck:2019nip},
the red and blue lines show the limits from simulated ground observation and PICO mission respectively,
and the green lines represent the prospective limits from a cosmic variance limited (CVL) mission observing in multipole range of $\ell \in [10,3000]$.

The effect of using different ILC methods on the constraints for DM parameters can be seen from Figure \ref{fig:pConstraints}. For ground observation, since the foreground residual provided by the two ILC methods is basically about the same level, so the final constraints on DM parameters from the two are closer to each other. For PICO, the advantages of NILC are more obvious. For the annihilation channel, the limits given by NILC is about 12\% ($\gamma\gamma$) and 8\% ($e^-e^+$)higher than that of HILC, and for decay chanels, the average sensitivity reduction is 15\% and 20\%. This result should be due to PICO's low noise level and multi-band can better play the advantages of NILC. In addition, due to the different foreground remaining levels caused by different foreground models used, the corresponding dark matter parameter constraints are also slightly different. The uncertainty provided by V2 model are slightly worse than those obtained by V1. However, since the foreground residual difference is not obvious, the influence on the result of parameter limitation is relatively small.

Our results show that the future observations can potentially give significantly improved limits on DM decay and annihilation compared to the current Planck experiment.
Taking an example of DM candidate with mass $m_\chi \sim 0.1 \GeV$, for an ideal case by using NILC method and simpler foreground V1, 
for the $\gamma \gamma$ annihilation channel,
ground-based CMB observations can push the constraint on annihilation cross-section to $\sv /m_\chi < 4.5 \times 10^{-28} \r{cm^3 GeV^{-1} s^{-1}}$,
which is about factor of 1.2 better than the Planck limit of $\sv /m_\chi< 5.6 \times 10 ^ {-28} \r{cm^3 GeV^{-1} s^{-1}}$,
while PICO can further improve $\sv /m_\chi$ constraint by about $3.1$ times.
For the decay channel $\chi\rightarrow\gamma\gamma$, 
future ground-based observations and PICO can potentially constrain $\Gamma_{\chi}$ to $\Gamma_\chi<  1.3 \times 10 ^ {-25} \r{s^{-1}}$ and $\Gamma_\chi< 3.7 \times 10 ^ {-26}\r{s^{-1}}$ respectively,
and a CVL experiment can further tighten the limits to $\Gamma_\chi< 3.3 \times 10 ^ {-26} \r{s^{-1}}$.
The specific limits for dark matter annihilation or decay to $\r{e}^{\pm}$ are shown in the right panels.


Such improvements demonstrate that indirect WIMP search promises of upcoming experiments persist after a proper treatment of the heavy CMB foreground. 
When compared to the standard thermal WIMP cross-section of $\vev{\sigma v}\sim 3 \times 10^{-26} \r{cm^3s^{-1}}$ in Figure~\ref{fig:pConstraints}, still consider using the foreground model V1 and removing the foreground with the NILC method,
the cross-over points indicate the exclusion of a vanilla thermal WIMP below 35 (83) and 44 (96) GeV in $\gamma\gamma$ and $e^+e^-$ annihilation channels with future ground-based (space-borne) programs, 
respectively. 
Here we only showed the DM limits from annihilation and decay channels into $e^+e^-$ and $\gamma\gamma$. 
For a more generic process into a pair of SM particle, 
the corresponding limits can be derived from folding in the final-state energy fraction in the form of electrons and photons, 
leading to an order ${\cal O}(0.1)$ factor for limits in $WW, ZZ, t\bar{t}$ and hadronic channels.

%% file: conclusion.tex
To summarize, 
in this work we perform a full foreground analysis on using the CMB temperature and polarization maps for constraining dark matter induced cosmic energy injection. 
We construct realistic microwave sky simulations to produce sky-maps that contain the CMB signal, 
synchrotron and dust foreground emissions.
We use the Planck Sky Model for the polarized foreground maps. 
Considering future ground-based and space-borne CMB experimental setups like some ground telecopes and PICO, 
the temperature map foreground far exceeds the level of instrumental noise on some $\ell$ scales. 
In comparison the polarization foreground is less severe, 
and it can be sub-dominant to instrumental noise in the relatively high $\ell$ range. 
As the polarization foreground can be better controlled, 
foreground contamination is less severe for polarization-dominated sensitivity, 
and significantly improved results can be obtained after proper foreground subtraction.


Early ionization effects near the recombination era can leave distinctive signatures on CMB polarization,
and we show here that CMB constraints on DM energy injection rely mostly on the polarization data.
For the ground observations,
we show that compared to the limits set by TT data alone,
combining TT, TE and EE datasets can improve DM constraints by about two orders of magnitude.
We report that such relative advantage with polarization data is still present with a careful ILC treatment of TT and EE foreground contamination.

After the inclusion of polarization foregrounds, 
future CMB polarization's prospective improvement on DM annihilation and decay are very significant.  
Comparing with statistics(noise)-only analyses, 
the presence of CMB foreground residual weakens the DM constraint by a factor of 7\%-23\% for different terrestrial and space-borne experimental configurations. 
After foreground residual removal,
the future space-borne experiment's sensitivity to DM decay lifetime can improve by around one order of magnitude over the existing Planck limits. 
Future ground-based experiment is also expected to score a comparably smaller improvement. 
For ground observation and PICO,
the respective exclusion limit for thermal WIMP annihilation can be extend to $m_\chi>44$ GeV and $m_\chi>96$ GeV in the $e^+e^-$ channel, 
or $m_\chi>35$ GeV and $m_\chi>83$ GeV in the $\gamma\gamma$ channel. 
Therefore, 
we expect the upcoming CMB experiments to play an important role in the quest of dark matter search, 
among many other scientific goals. 

%% file: main.bbl
\providecommand{\href}[2]{#2}\begingroup\raggedright\begin{thebibliography}{10}

\bibitem{Rubin:1970zza}
V.~C. Rubin and W.~K. Ford, Jr., \emph{{Rotation of the Andromeda Nebula from a
  Spectroscopic Survey of Emission Regions}},
  \href{https://doi.org/10.1086/150317}{\emph{Astrophys. J.} {\bfseries 159}
  (1970) 379--403}.

\bibitem{Rubin:1980zd}
V.~C. Rubin, N.~Thonnard and W.~K. Ford, Jr., \emph{{Rotational properties of
  21 SC galaxies with a large range of luminosities and radii, from NGC 4605 /R
  = 4kpc/ to UGC 2885 /R = 122 kpc/}},
  \href{https://doi.org/10.1086/158003}{\emph{Astrophys. J.} {\bfseries 238}
  (1980) 471}.

\bibitem{vanAlbada:1984js}
T.~S. van Albada, J.~N. Bahcall, K.~Begeman and R.~Sancisi, \emph{{The
  Distribution of Dark Matter in the Spiral Galaxy {NGC}-3198}},
  \href{https://doi.org/10.1086/163375}{\emph{Astrophys. J.} {\bfseries 295}
  (1985) 305--313}.

\bibitem{Clowe_2006}
D.~Clowe, M.~Bradač, A.~H. Gonzalez, M.~Markevitch, S.~W. Randall, C.~Jones
  et~al., \emph{A direct empirical proof of the existence of dark matter*},
  \href{https://doi.org/10.1086/508162}{\emph{The Astrophysical Journal}
  {\bfseries 648} (aug, 2006) L109}.

\bibitem{Clowe_2004}
D.~Clowe, A.~Gonzalez and M.~Markevitch, \emph{Weak-lensing mass reconstruction
  of the interacting cluster 1e 0657–558: Direct evidence for the existence
  of dark matter*}, \href{https://doi.org/10.1086/381970}{\emph{The
  Astrophysical Journal} {\bfseries 604} (apr, 2004) 596}.

\bibitem{Markevitch_2002}
M.~Markevitch, A.~H. Gonzalez, L.~David, A.~Vikhlinin, S.~Murray, W.~Forman
  et~al., \emph{A textbook example of a bow shock in the merging galaxy cluster
  1e 0657–56}, \href{https://doi.org/10.1086/339619}{\emph{The Astrophysical
  Journal} {\bfseries 567} (feb, 2002) L27}.

\bibitem{Planck:2018vyg}
{\scshape Planck} collaboration, N.~Aghanim et~al., \emph{{Planck 2018 results.
  VI. Cosmological parameters}},
  \href{https://doi.org/10.1051/0004-6361/201833910}{\emph{Astron. Astrophys.}
  {\bfseries 641} (2020) A6},
  [\href{https://arxiv.org/abs/1807.06209}{{\ttfamily 1807.06209}}].

\bibitem{Albouy:2022cin}
G.~Albouy et~al., \emph{{Theory, phenomenology, and experimental avenues for
  dark showers: a Snowmass 2021 report}},
  \href{https://doi.org/10.1140/epjc/s10052-022-11048-8}{\emph{Eur. Phys. J. C}
  {\bfseries 82} (2022) 1132},
  [\href{https://arxiv.org/abs/2203.09503}{{\ttfamily 2203.09503}}].

\bibitem{PhysRevLett.50.1419}
H.~Goldberg, \emph{Constraint on the photino mass from cosmology},
  \href{https://doi.org/10.1103/PhysRevLett.50.1419}{\emph{Phys. Rev. Lett.}
  {\bfseries 50} (May, 1983) 1419--1422}.

\bibitem{PhysRevLett.89.211301}
H.-C. Cheng, J.~L. Feng and K.~T. Matchev, \emph{Kaluza-klein dark matter},
  \href{https://doi.org/10.1103/PhysRevLett.89.211301}{\emph{Phys. Rev. Lett.}
  {\bfseries 89} (Oct, 2002) 211301}.

\bibitem{SERVANT2003391}
G.~Servant and T.~M. Tait, \emph{Is the lightest kaluza–klein particle a
  viable dark matter candidate?},
  \href{https://doi.org/https://doi.org/10.1016/S0550-3213(02)01012-X}{\emph{Nuclear
  Physics B} {\bfseries 650} (2003) 391--419}.

\bibitem{ELLIS1984453}
J.~Ellis, J.~Hagelin, D.~Nanopoulos, K.~Olive and M.~Srednicki,
  \emph{Supersymmetric relics from the big bang},
  \href{https://doi.org/https://doi.org/10.1016/0550-3213(84)90461-9}{\emph{Nuclear
  Physics B} {\bfseries 238} (1984) 453--476}.

\bibitem{Bertone:2018krk}
G.~Bertone and T.~Tait, M.~P., \emph{{A new era in the search for dark
  matter}}, \href{https://doi.org/10.1038/s41586-018-0542-z}{\emph{Nature}
  {\bfseries 562} (2018) 51--56},
  [\href{https://arxiv.org/abs/1810.01668}{{\ttfamily 1810.01668}}].

\bibitem{Shi:1998km}
X.-D. Shi and G.~M. Fuller, \emph{{A New dark matter candidate: Nonthermal
  sterile neutrinos}},
  \href{https://doi.org/10.1103/PhysRevLett.82.2832}{\emph{Phys. Rev. Lett.}
  {\bfseries 82} (1999) 2832--2835},
  [\href{https://arxiv.org/abs/astro-ph/9810076}{{\ttfamily
  astro-ph/9810076}}].

\bibitem{Laine:2008pg}
M.~Laine and M.~Shaposhnikov, \emph{{Sterile neutrino dark matter as a
  consequence of nuMSM-induced lepton asymmetry}},
  \href{https://doi.org/10.1088/1475-7516/2008/06/031}{\emph{JCAP} {\bfseries
  06} (2008) 031}, [\href{https://arxiv.org/abs/0804.4543}{{\ttfamily
  0804.4543}}].

\bibitem{Peccei:1977hh}
R.~D. Peccei and H.~R. Quinn, \emph{{CP Conservation in the Presence of
  Instantons}}, \href{https://doi.org/10.1103/PhysRevLett.38.1440}{\emph{Phys.
  Rev. Lett.} {\bfseries 38} (1977) 1440--1443}.

\bibitem{Sikivie:1982qv}
P.~Sikivie, \emph{{Of Axions, Domain Walls and the Early Universe}},
  \href{https://doi.org/10.1103/PhysRevLett.48.1156}{\emph{Phys. Rev. Lett.}
  {\bfseries 48} (1982) 1156--1159}.

\bibitem{Preskill:1982cy}
J.~Preskill, M.~B. Wise and F.~Wilczek, \emph{{Cosmology of the Invisible
  Axion}}, \href{https://doi.org/10.1016/0370-2693(83)90637-8}{\emph{Phys.
  Lett. B} {\bfseries 120} (1983) 127--132}.

\bibitem{Padmanabhan:2005es}
N.~Padmanabhan and D.~P. Finkbeiner, \emph{{Detecting dark matter annihilation
  with CMB polarization: Signatures and experimental prospects}},
  \href{https://doi.org/10.1103/PhysRevD.72.023508}{\emph{Phys. Rev. D}
  {\bfseries 72} (2005) 023508},
  [\href{https://arxiv.org/abs/astro-ph/0503486}{{\ttfamily
  astro-ph/0503486}}].

\bibitem{Zhang:2007zzh}
L.~Zhang, X.~Chen, M.~Kamionkowski, Z.-g. Si and Z.~Zheng, \emph{{Constraints
  on radiative dark-matter decay from the cosmic microwave background}},
  \href{https://doi.org/10.1103/PhysRevD.76.061301}{\emph{Phys. Rev. D}
  {\bfseries 76} (2007) 061301},
  [\href{https://arxiv.org/abs/0704.2444}{{\ttfamily 0704.2444}}].

\bibitem{Madhavacheril:2013cna}
M.~S. Madhavacheril, N.~Sehgal and T.~R. Slatyer, \emph{{Current Dark Matter
  Annihilation Constraints from CMB and Low-Redshift Data}},
  \href{https://doi.org/10.1103/PhysRevD.89.103508}{\emph{Phys. Rev. D}
  {\bfseries 89} (2014) 103508},
  [\href{https://arxiv.org/abs/1310.3815}{{\ttfamily 1310.3815}}].

\bibitem{Slatyer:2015kla}
T.~R. Slatyer, \emph{{Indirect Dark Matter Signatures in the Cosmic Dark Ages
  II. Ionization, Heating and Photon Production from Arbitrary Energy
  Injections}}, \href{https://doi.org/10.1103/PhysRevD.93.023521}{\emph{Phys.
  Rev. D} {\bfseries 93} (2016) 023521},
  [\href{https://arxiv.org/abs/1506.03812}{{\ttfamily 1506.03812}}].

\bibitem{PhysRevD.85.043522}
D.~P. Finkbeiner, S.~Galli, T.~Lin and T.~R. Slatyer, \emph{Searching for dark
  matter in the cmb: A compact parametrization of energy injection from new
  physics}, \href{https://doi.org/10.1103/PhysRevD.85.043522}{\emph{Phys. Rev.
  D} {\bfseries 85} (Feb, 2012) 043522}.

\bibitem{Fermi-LAT:2011vow}
{\scshape Fermi-LAT} collaboration, M.~Ackermann et~al., \emph{{Constraining
  Dark Matter Models from a Combined Analysis of Milky Way Satellites with the
  Fermi Large Area Telescope}},
  \href{https://doi.org/10.1103/PhysRevLett.107.241302}{\emph{Phys. Rev. Lett.}
  {\bfseries 107} (2011) 241302},
  [\href{https://arxiv.org/abs/1108.3546}{{\ttfamily 1108.3546}}].

\bibitem{CMB-S4:2016ple}
{\scshape CMB-S4} collaboration, K.~N. Abazajian et~al., \emph{{CMB-S4 Science
  Book, First Edition}},  \href{https://arxiv.org/abs/1610.02743}{{\ttfamily
  1610.02743}}.

\bibitem{Abazajian:2019eic}
K.~Abazajian et~al., \emph{{CMB-S4 Science Case, Reference Design, and Project
  Plan}},  \href{https://arxiv.org/abs/1907.04473}{{\ttfamily 1907.04473}}.

\bibitem{SimonsObservatory:2019qwx}
{\scshape Simons Observatory} collaboration, M.~H. Abitbol et~al., \emph{{The
  Simons Observatory: Astro2020 Decadal Project Whitepaper}}, {\emph{Bull. Am.
  Astron. Soc.} {\bfseries 51} (2019) 147},
  [\href{https://arxiv.org/abs/1907.08284}{{\ttfamily 1907.08284}}].

\bibitem{Li:2017drr}
H.~Li et~al., \emph{{Probing Primordial Gravitational Waves: Ali CMB
  Polarization Telescope}},
  \href{https://doi.org/10.1093/nsr/nwy019}{\emph{Natl. Sci. Rev.} {\bfseries
  6} (2019) 145--154}, [\href{https://arxiv.org/abs/1710.03047}{{\ttfamily
  1710.03047}}].

\bibitem{Li:2018rwc}
H.~Li, S.-Y. Li, Y.~Liu, Y.-P. Li and X.~Zhang, \emph{{Tibet\textquoteright{}s
  window on primordial gravitational waves}},
  \href{https://doi.org/10.1038/s41550-017-0373-0}{\emph{Nature Astron.}
  {\bfseries 2} (2018) 104--106},
  [\href{https://arxiv.org/abs/1802.08455}{{\ttfamily 1802.08455}}].

\bibitem{Cang:2020exa}
J.~Cang, Y.~Gao and Y.-Z. Ma, \emph{{Probing dark matter with future CMB
  measurements}},
  \href{https://doi.org/10.1103/PhysRevD.102.103005}{\emph{Phys. Rev. D}
  {\bfseries 102} (2020) 103005},
  [\href{https://arxiv.org/abs/2002.03380}{{\ttfamily 2002.03380}}].

\bibitem{2003MNRAS.345.1101M}
E.~{Mart{\'\i}nez-Gonz{\'a}lez}, J.~M. {Diego}, P.~{Vielva} and J.~{Silk},
  \emph{{Cosmic microwave background power spectrum estimation and map
  reconstruction with the expectation-maximization algorithm}},
  \href{https://doi.org/10.1046/j.1365-2966.2003.06885.x}{\emph{\mnras}
  {\bfseries 345} (Nov., 2003) 1101--1109},
  [\href{https://arxiv.org/abs/astro-ph/0302094}{{\ttfamily
  astro-ph/0302094}}].

\bibitem{2008A&A...491..597L}
{{S. M. Leach}, {J. F. Cardoso}, {C. Baccigalupi}, et \emph{al.}},
  \emph{{Component separation methods for the PLANCK mission}},
  \href{https://doi.org/10.1051/0004-6361:200810116}{\emph{\aap} {\bfseries
  491} (Nov., 2008) 597--615},
  [\href{https://arxiv.org/abs/0805.0269}{{\ttfamily 0805.0269}}].

\bibitem{2012MNRAS.420.2162F}
R.~{Fern{\'a}ndez-Cobos}, P.~{Vielva}, R.~B. {Barreiro} and
  E.~{Mart{\'\i}nez-Gonz{\'a}lez}, \emph{{Multiresolution internal template
  cleaning: an application to the Wilkinson Microwave Anisotropy Probe 7-yr
  polarization data}},
  \href{https://doi.org/10.1111/j.1365-2966.2011.20182.x}{\emph{\mnras}
  {\bfseries 420} (Mar., 2012) 2162--2169},
  [\href{https://arxiv.org/abs/1106.2016}{{\ttfamily 1106.2016}}].

\bibitem{2019PTEP.2019c3E01I}
K.~{Ichiki}, H.~{Kanai}, N.~{Katayama} and E.~{Komatsu}, \emph{{Delta-map
  method of removing CMB foregrounds with spatially varying spectra}},
  \href{https://doi.org/10.1093/ptep/ptz009}{\emph{Progress of Theoretical and
  Experimental Physics} {\bfseries 2019} (Mar., 2019) 033E01},
  [\href{https://arxiv.org/abs/1811.03886}{{\ttfamily 1811.03886}}].

\bibitem{2006ApJ...641..665E}
H.~K. {Eriksen}, C.~{Dickinson}, C.~R. {Lawrence}, C.~{Baccigalupi}, A.~J.
  {Banday}, K.~M. {G{\'o}rski} et~al., \emph{{Cosmic Microwave Background
  Component Separation by Parameter Estimation}},
  \href{https://doi.org/10.1086/500499}{\emph{\apj} {\bfseries 641} (Apr.,
  2006) 665--682}, [\href{https://arxiv.org/abs/astro-ph/0508268}{{\ttfamily
  astro-ph/0508268}}].

\bibitem{fgb-Stompor:2008sf}
R.~Stompor, S.~M. Leach, F.~Stivoli and C.~Baccigalupi, \emph{{Maximum
  Likelihood algorithm for parametric component separation in CMB
  experiments}},
  \href{https://doi.org/10.1111/j.1365-2966.2008.14023.x}{\emph{Mon. Not. Roy.
  Astron. Soc.} {\bfseries 392} (2009) 216},
  [\href{https://arxiv.org/abs/0804.2645}{{\ttfamily 0804.2645}}].

\bibitem{2019MNRAS.484.1616Z}
P.~{Zhang}, J.~{Zhang} and L.~{Zhang}, \emph{{ABS: an analytical method of
  blind separation of CMB from foregrounds}},
  \href{https://doi.org/10.1093/mnras/stz091}{\emph{\mnras} {\bfseries 484}
  (Apr., 2019) 1616--1626}.

\bibitem{2013MNRAS.435...18B}
S.~{Basak} and J.~{Delabrouille}, \emph{{A needlet ILC analysis of WMAP 9-year
  polarization data: CMB polarization power spectra}},
  \href{https://doi.org/10.1093/mnras/stt1158}{\emph{\mnras} {\bfseries 435}
  (Oct., 2013) 18--29}, [\href{https://arxiv.org/abs/1204.0292}{{\ttfamily
  1204.0292}}].

\bibitem{2003ApJS..148...97B}
{{C. L. Bennett}, {R. S. Hill}, {G. Hinshaw}, et \emph{al.}}, \emph{{First-Year
  Wilkinson Microwave Anisotropy Probe (WMAP) Observations: Foreground
  Emission}}, \href{https://doi.org/10.1086/377252}{\emph{\apjs} {\bfseries
  148} (Sept., 2003) 97--117},
  [\href{https://arxiv.org/abs/astro-ph/0302208}{{\ttfamily
  astro-ph/0302208}}].

\bibitem{2013ApJS..208...20B}
C.~L. {Bennett}, D.~{Larson}, J.~L. {Weiland}, N.~{Jarosik}, G.~{Hinshaw},
  N.~{Odegard} et~al., \emph{{Nine-year Wilkinson Microwave Anisotropy Probe
  (WMAP) Observations: Final Maps and Results}},
  \href{https://doi.org/10.1088/0067-0049/208/2/20}{\emph{\apjs} {\bfseries
  208} (Oct., 2013) 20}, [\href{https://arxiv.org/abs/1212.5225}{{\ttfamily
  1212.5225}}].

\bibitem{2016A&A...594A...9P}
{Planck Collaboration}, R.~{Adam}, P.~A.~R. {Ade}, N.~{Aghanim}, M.~{Arnaud},
  M.~{Ashdown} et~al., \emph{{Planck 2015 results. IX. Diffuse component
  separation: CMB maps}},
  \href{https://doi.org/10.1051/0004-6361/201525936}{\emph{\aap} {\bfseries
  594} (Sept., 2016) A9}, [\href{https://arxiv.org/abs/1502.05956}{{\ttfamily
  1502.05956}}].

\bibitem{Liu:2016cnk}
H.~Liu, T.~R. Slatyer and J.~Zavala, \emph{{Contributions to cosmic
  reionization from dark matter annihilation and decay}},
  \href{https://doi.org/10.1103/PhysRevD.94.063507}{\emph{Phys. Rev. D}
  {\bfseries 94} (2016) 063507},
  [\href{https://arxiv.org/abs/1604.02457}{{\ttfamily 1604.02457}}].

\bibitem{Cang:2021owu}
J.~Cang, Y.~Gao and Y.-Z. Ma, \emph{{21-cm constraints on spinning primordial
  black holes}},
  \href{https://doi.org/10.1088/1475-7516/2022/03/012}{\emph{JCAP} {\bfseries
  03} (2022) 012}, [\href{https://arxiv.org/abs/2108.13256}{{\ttfamily
  2108.13256}}].

\bibitem{Slatyer:2016qyl}
T.~R. Slatyer and C.-L. Wu, \emph{{General Constraints on Dark Matter Decay
  from the Cosmic Microwave Background}},
  \href{https://doi.org/10.1103/PhysRevD.95.023010}{\emph{Phys. Rev. D}
  {\bfseries 95} (2017) 023010},
  [\href{https://arxiv.org/abs/1610.06933}{{\ttfamily 1610.06933}}].

\bibitem{Ali-Haimoud:2010hou}
Y.~Ali-Haimoud and C.~M. Hirata, \emph{{HyRec: A fast and highly accurate
  primordial hydrogen and helium recombination code}},
  \href{https://doi.org/10.1103/PhysRevD.83.043513}{\emph{Phys. Rev. D}
  {\bfseries 83} (2011) 043513},
  [\href{https://arxiv.org/abs/1011.3758}{{\ttfamily 1011.3758}}].

\bibitem{Seager:1999bc}
S.~Seager, D.~D. Sasselov and D.~Scott, \emph{{A new calculation of the
  recombination epoch}}, \href{https://doi.org/10.1086/312250}{\emph{Astrophys.
  J. Lett.} {\bfseries 523} (1999) L1--L5},
  [\href{https://arxiv.org/abs/astro-ph/9909275}{{\ttfamily
  astro-ph/9909275}}].

\bibitem{Lewis:1999bs}
A.~Lewis, A.~Challinor and A.~Lasenby, \emph{{Efficient computation of CMB
  anisotropies in closed FRW models}},
  \href{https://doi.org/10.1086/309179}{\emph{Astrophys. J.} {\bfseries 538}
  (2000) 473--476}, [\href{https://arxiv.org/abs/astro-ph/9911177}{{\ttfamily
  astro-ph/9911177}}].

\bibitem{Lopez-Honorez:2016sur}
L.~Lopez-Honorez, O.~Mena, A.~Molin\'e, S.~Palomares-Ruiz and A.~C. Vincent,
  \emph{{The 21 cm signal and the interplay between dark matter annihilations
  and astrophysical processes}},
  \href{https://doi.org/10.1088/1475-7516/2016/08/004}{\emph{JCAP} {\bfseries
  08} (2016) 004}, [\href{https://arxiv.org/abs/1603.06795}{{\ttfamily
  1603.06795}}].

\bibitem{Clark:2018ghm}
S.~Clark, B.~Dutta, Y.~Gao, Y.-Z. Ma and L.~E. Strigari, \emph{{21 cm limits on
  decaying dark matter and primordial black holes}},
  \href{https://doi.org/10.1103/PhysRevD.98.043006}{\emph{Phys. Rev. D}
  {\bfseries 98} (2018) 043006},
  [\href{https://arxiv.org/abs/1803.09390}{{\ttfamily 1803.09390}}].

\bibitem{Lewis:2002ah}
A.~Lewis and S.~Bridle, \emph{{Cosmological parameters from CMB and other data:
  A Monte Carlo approach}},
  \href{https://doi.org/10.1103/PhysRevD.66.103511}{\emph{Phys. Rev. D}
  {\bfseries 66} (2002) 103511},
  [\href{https://arxiv.org/abs/astro-ph/0205436}{{\ttfamily
  astro-ph/0205436}}].

\bibitem{Lewis:2013hha}
A.~Lewis, \emph{{Efficient sampling of fast and slow cosmological parameters}},
  \href{https://doi.org/10.1103/PhysRevD.87.103529}{\emph{Phys. Rev. D}
  {\bfseries 87} (2013) 103529},
  [\href{https://arxiv.org/abs/1304.4473}{{\ttfamily 1304.4473}}].

\bibitem{2016A&A...594A..10P}
{Planck Collaboration}, R.~{Adam}, P.~A.~R. {Ade}, N.~{Aghanim}, M.~I.~R.
  {Alves}, M.~{Arnaud} et~al., \emph{{Planck 2015 results. X. Diffuse component
  separation: Foreground maps}},
  \href{https://doi.org/10.1051/0004-6361/201525967}{\emph{\aap} {\bfseries
  594} (Sept., 2016) A10}, [\href{https://arxiv.org/abs/1502.01588}{{\ttfamily
  1502.01588}}].

\bibitem{2013A&A...553A..96D}
{{J. Delabrouille}, {M. Betoule}, {J. B. Melin}, et \emph{al.}}, \emph{{The
  pre-launch Planck Sky Model: a model of sky emission at submillimetre to
  centimetre wavelengths}},
  \href{https://doi.org/10.1051/0004-6361/201220019}{\emph{\aap} {\bfseries
  553} (May, 2013) A96}, [\href{https://arxiv.org/abs/1207.3675}{{\ttfamily
  1207.3675}}].

\bibitem{2013MNRAS.436.2127O}
E.~{Orlando} and A.~{Strong}, \emph{{Galactic synchrotron emission with cosmic
  ray propagation models}},
  \href{https://doi.org/10.1093/mnras/stt1718}{\emph{\mnras} {\bfseries 436}
  (Dec., 2013) 2127--2142}, [\href{https://arxiv.org/abs/1309.2947}{{\ttfamily
  1309.2947}}].

\bibitem{1982A&AS...47....1H}
C.~G.~T. {Haslam}, C.~J. {Salter}, H.~{Stoffel} and W.~E. {Wilson}, \emph{{A
  408-MHZ All-Sky Continuum Survey. II. The Atlas of Contour Maps}},
  {\emph{\aaps} {\bfseries 47} (Jan., 1982) 1}.

\bibitem{2015MNRAS.451.4311R}
M.~{Remazeilles}, C.~{Dickinson}, A.~J. {Banday}, M.~A. {Bigot-Sazy} and
  T.~{Ghosh}, \emph{{An improved source-subtracted and destriped 408-MHz
  all-sky map}}, \href{https://doi.org/10.1093/mnras/stv1274}{\emph{\mnras}
  {\bfseries 451} (Aug., 2015) 4311--4327},
  [\href{https://arxiv.org/abs/1411.3628}{{\ttfamily 1411.3628}}].

\bibitem{2020A&A...641A...4P}
{Planck Collaboration}, Y.~{Akrami}, M.~{Ashdown}, J.~{Aumont},
  C.~{Baccigalupi}, M.~{Ballardini} et~al., \emph{{Planck 2018 results. IV.
  Diffuse component separation}},
  \href{https://doi.org/10.1051/0004-6361/201833881}{\emph{\aap} {\bfseries
  641} (Sept., 2020) A4}, [\href{https://arxiv.org/abs/1807.06208}{{\ttfamily
  1807.06208}}].

\bibitem{2014A&A...571A..11P}
{Planck Collaboration}, A.~{Abergel}, P.~A.~R. {Ade}, N.~{Aghanim}, M.~I.~R.
  {Alves}, G.~{Aniano} et~al., \emph{{Planck 2013 results. XI. All-sky model of
  thermal dust emission}},
  \href{https://doi.org/10.1051/0004-6361/201323195}{\emph{\aap} {\bfseries
  571} (Nov., 2014) A11}, [\href{https://arxiv.org/abs/1312.1300}{{\ttfamily
  1312.1300}}].

\bibitem{2016A&A...596A.109P}
{Planck Collaboration}, N.~{Aghanim}, M.~{Ashdown}, J.~{Aumont},
  C.~{Baccigalupi}, M.~{Ballardini} et~al., \emph{{Planck intermediate results.
  XLVIII. Disentangling Galactic dust emission and cosmic infrared background
  anisotropies}},
  \href{https://doi.org/10.1051/0004-6361/201629022}{\emph{\aap} {\bfseries
  596} (Dec., 2016) A109}, [\href{https://arxiv.org/abs/1605.09387}{{\ttfamily
  1605.09387}}].

\bibitem{2019JCAP...02..056A}
P.~{Ade}, J.~{Aguirre}, Z.~{Ahmed}, S.~{Aiola}, A.~{Ali}, D.~{Alonso} et~al.,
  \emph{{The Simons Observatory: science goals and forecasts}},
  \href{https://doi.org/10.1088/1475-7516/2019/02/056}{\emph{JCAP} {\bfseries
  2019} (Feb, 2019) 056}, [\href{https://arxiv.org/abs/1808.07445}{{\ttfamily
  1808.07445}}].

\bibitem{2019arXiv190210541H}
S.~{Hanany}, M.~{Alvarez}, E.~{Artis}, P.~{Ashton}, J.~{Aumont}, R.~{Aurlien}
  et~al., \emph{{PICO: Probe of Inflation and Cosmic Origins}}, {\emph{arXiv
  e-prints} (Feb., 2019) arXiv:1902.10541},
  [\href{https://arxiv.org/abs/1902.10541}{{\ttfamily 1902.10541}}].

\bibitem{Aurlien:2022tlp}
R.~Aurlien et~al., \emph{{Foreground Separation and Constraints on Primordial
  Gravitational Waves with the PICO Space Mission}},
  \href{https://arxiv.org/abs/2211.14342}{{\ttfamily 2211.14342}}.

\bibitem{Errard:2015cxa}
J.~Errard, S.~M. Feeney, H.~V. Peiris and A.~H. Jaffe, \emph{{Robust forecasts
  on fundamental physics from the foreground-obscured, gravitationally-lensed
  CMB polarization}},
  \href{https://doi.org/10.1088/1475-7516/2016/03/052}{\emph{JCAP} {\bfseries
  03} (2016) 052}, [\href{https://arxiv.org/abs/1509.06770}{{\ttfamily
  1509.06770}}].

\bibitem{2020A&A...643A..42P}
{Planck Collaboration}, Y.~{Akrami}, K.~J. {Andersen}, M.~{Ashdown},
  C.~{Baccigalupi}, M.~{Ballardini} et~al., \emph{{Planck intermediate results.
  LVII. Joint Planck LFI and HFI data processing}},
  \href{https://doi.org/10.1051/0004-6361/202038073}{\emph{\aap} {\bfseries
  643} (Nov., 2020) A42}, [\href{https://arxiv.org/abs/2007.04997}{{\ttfamily
  2007.04997}}].

\bibitem{Tegmark:2003ve}
M.~Tegmark, A.~de~Oliveira-Costa and A.~Hamilton, \emph{{A high resolution
  foreground cleaned CMB map from WMAP}},
  \href{https://doi.org/10.1103/PhysRevD.68.123523}{\emph{Phys. Rev. D}
  {\bfseries 68} (2003) 123523},
  [\href{https://arxiv.org/abs/astro-ph/0302496}{{\ttfamily
  astro-ph/0302496}}].

\bibitem{Hamimeche:2008ai}
S.~Hamimeche and A.~Lewis, \emph{{Likelihood Analysis of CMB Temperature and
  Polarization Power Spectra}},
  \href{https://doi.org/10.1103/PhysRevD.77.103013}{\emph{Phys. Rev. D}
  {\bfseries 77} (2008) 103013},
  [\href{https://arxiv.org/abs/0801.0554}{{\ttfamily 0801.0554}}].

\bibitem{Foreman-Mackey:2012any}
D.~Foreman-Mackey, D.~W. Hogg, D.~Lang and J.~Goodman, \emph{{emcee: The MCMC
  Hammer}}, \href{https://doi.org/10.1086/670067}{\emph{Publ. Astron. Soc.
  Pac.} {\bfseries 125} (2013) 306--312},
  [\href{https://arxiv.org/abs/1202.3665}{{\ttfamily 1202.3665}}].

\bibitem{Planck:2019nip}
{\scshape Planck} collaboration, N.~Aghanim et~al., \emph{{Planck 2018 results.
  V. CMB power spectra and likelihoods}},
  \href{https://doi.org/10.1051/0004-6361/201936386}{\emph{Astron. Astrophys.}
  {\bfseries 641} (2020) A5},
  [\href{https://arxiv.org/abs/1907.12875}{{\ttfamily 1907.12875}}].

\end{thebibliography}\endgroup
